\documentclass[english,10pt,aps,prx,twocolumn,superscriptaddress,amsmath,amssymb,floatfix,noeprint]{revtex4-2}
\usepackage{graphicx}
\usepackage{physics}
\usepackage{xcolor}
\usepackage[normalem]{ulem} 
\usepackage{bbm}
\usepackage[colorlinks=true,citecolor=blue,linkcolor=blue,urlcolor=blue]{hyperref}
\usepackage{comment}

\begin{document}

\title{\texorpdfstring{Probing entanglement scaling across a quantum phase transition\\ on a quantum computer}{Probing entanglement scaling across a quantum phase transition on a quantum computer}}

\author{Qiang Miao}
  \affiliation{Duke Quantum Center, Duke University, Durham, North Carolina 27701, USA}
  
\author{Tianyi Wang}
  \affiliation{Duke Quantum Center, Duke University, Durham, North Carolina 27701, USA}
  \affiliation{Department of Physics, Duke University, Durham, North Carolina 27708, USA}

\author{Kenneth R. Brown}
  \affiliation{Duke Quantum Center, Duke University, Durham, North Carolina 27701, USA}
  \affiliation{Department of Physics, Duke University, Durham, North Carolina 27708, USA}
  \affiliation{Department of Electrical and Computer Engineering, Duke University, Durham, North Carolina 27708, USA}
  \affiliation{Department of Chemistry, Duke University, Durham, North Carolina 27708, USA}
  
\author{Thomas Barthel}
  \email{Corresponding author, thomas.barthel@duke.edu}
  \affiliation{Duke Quantum Center, Duke University, Durham, North Carolina 27701, USA}
  \affiliation{Department of Physics, Duke University, Durham, North Carolina 27708, USA}
  \affiliation{National Quantum Laboratory, University of Maryland, College Park, MD 20742, USA}  
  \affiliation{Department of Physics, University of Maryland, College Park, MD 20742, USA}

\author{Marko Cetina}
  \email{Corresponding author, marko.cetina@duke.edu}
  \affiliation{Duke Quantum Center, Duke University, Durham, North Carolina 27701, USA}
  \affiliation{Department of Physics, Duke University, Durham, North Carolina 27708, USA}
  \affiliation{Department of Electrical and Computer Engineering, Duke University, Durham, North Carolina 27708, USA}

\date{April 2026}

\begin{abstract}

The investigation of strongly-correlated quantum matter is difficult due to the curse of dimensionality and intricate entanglement structures. These challenges are particularly pronounced in the vicinity of continuous quantum phase transitions, where quantum fluctuations manifest across all length scales. While quantum simulators give controlled access to a number of strongly correlated systems, the study of critical phenomena has been hampered by finite-size effects arising from diverging correlation lengths. Moreover, the experimental investigation of entanglement in many-body systems has been hindered by limitations in measurement protocols. To address these challenges, we employ the multiscale entanglement renormalization ansatz (MERA) and implement a holographic scheme for subsystem tomography on a fully-connected trapped-ion quantum computer. Our method accurately represents infinite systems and long-range correlations with few qubits, facilitating the efficient extraction of observables and entanglement properties, even at criticality. We observe a quantum phase transition with spontaneous symmetry breaking and reveal the evolution of entanglement properties across the critical point. For the first time, we demonstrate log-law scaling of subsystem entanglement entropies at criticality on a digital quantum computer. This achievement highlights the potential of MERA for the investigation of strongly-correlated many-body systems on quantum computers.
\end{abstract}

\maketitle

\section{Introduction}
In physics and materials science, universality appears near continuous phase transitions, where long-range properties become independent of microscopic details, and the singular behavior of physical quantities such as susceptibilities and correlation lengths can be characterized by universal critical exponents. While classical phase transitions are driven by thermal fluctuations, quantum phase transitions occur at zero temperature when a system parameter is varied, causing the ground state to become highly entangled due to quantum fluctuations~\cite{Sondhi1997-69,Sachdev2011}. This groundstate non-locality presents challenges for theoretical and experimental analysis.

Rapid advancements on controlled quantum platforms have expanded the range of experimentally accessible quantum many-body systems on quantum simulators and computers~\cite{Bloch2007,Smith2019-5,Barratt2021-7,Meth2022-12,Dborin2022-13,Zhu2022-128,Tajik2023-120,Fang2025-390}. Previous studies~\cite{Islam2015-528,li2019-5,bergschneider2019-15,Brydges2019-364,Foss-Feig2022-128,tajik2023-19,Joshi2023-624,Karamlou2024-629} have made significant progress in the experimental characterization of entanglement. However, probing entanglement properties across quantum phase transitions remains difficult. Analog quantum simulators have limited control, restricting available models and measurements~\cite{Barthel2018-121}. Quantum simulations using quantum computers face challenges including:
(i) \textit{Limited number of qubits}: Quantum phase transitions formally occur only in infinite systems, and finite-size effects are pronounced in the vicinity of critical points. Preparing critical ground states directly requires more qubits than current quantum devices offer.
(ii) \textit{Exponential cost for probing subsystem entanglement}: Resolving detailed entanglement structures requires state tomography for large subsystems, which generally results in exponential costs.
(iii) \textit{Difficulties in state preparation}: Without deeper insight into the physics, the study of complex many-body ground states necessitates highly expressive quantum circuits with many gates, leading to low trainability~\cite{McClean2018-9,Cerezo2021-12} and significant error accumulation.

We overcome these challenges for condensed-matter ground states using highly structured quantum circuits on a digital quantum computer. The circuits prepare states corresponding to hierarchical tensor networks---the multiscale entanglement renormalization ansatz (MERA)~\cite{Vidal2007-99,Vidal2008-101,Evenbly2009-79}. This ansatz can represent infinite systems and accurately encodes long-range correlations, even at critical points. The causal structure of MERA, inspired by the real-space renormalization group, drastically reduces circuit depths and the number of measured qubits, both scaling logarithmically with system size~\cite{Miao2023-5,Miao2023_03,Haghshenas2022-12,Haghshenas2024-133}. This allows efficient extraction of local observables, subsystem entanglement, and entanglement spectra. Additionally, MERA is robust to noise~\cite{Kim2017_11} and not affected by barren plateaus~\cite{Barthel2023_03,Miao2024-109}.

In this work, we study quantum many-body ground states on a digital quantum computer based on a chain of ytterbium ions. We demonstrate a quantum phase transition characterized by spontaneous symmetry breaking in the thermodynamic limit, using 12 or fewer qubits. We show the universal scaling of relevant physical quantities near the transition and observe significant changes in the entanglement structure across the critical point. At the critical point, we find that the subsystem entanglement entropies follow a universal log-area law, and the gap of the entanglement Hamiltonian closes as the subsystem size increases. In non-critical regimes, subsystem entanglement entropy follows an area law~\cite{Eisert2008,Latorre2009,Laflorencie2016-646}. The measured entanglement Hamiltonian remains gapped with increasing subsystem size and accurately reproduces expected symmetry properties.

\begin{figure}
    \includegraphics[width=1.0\columnwidth]{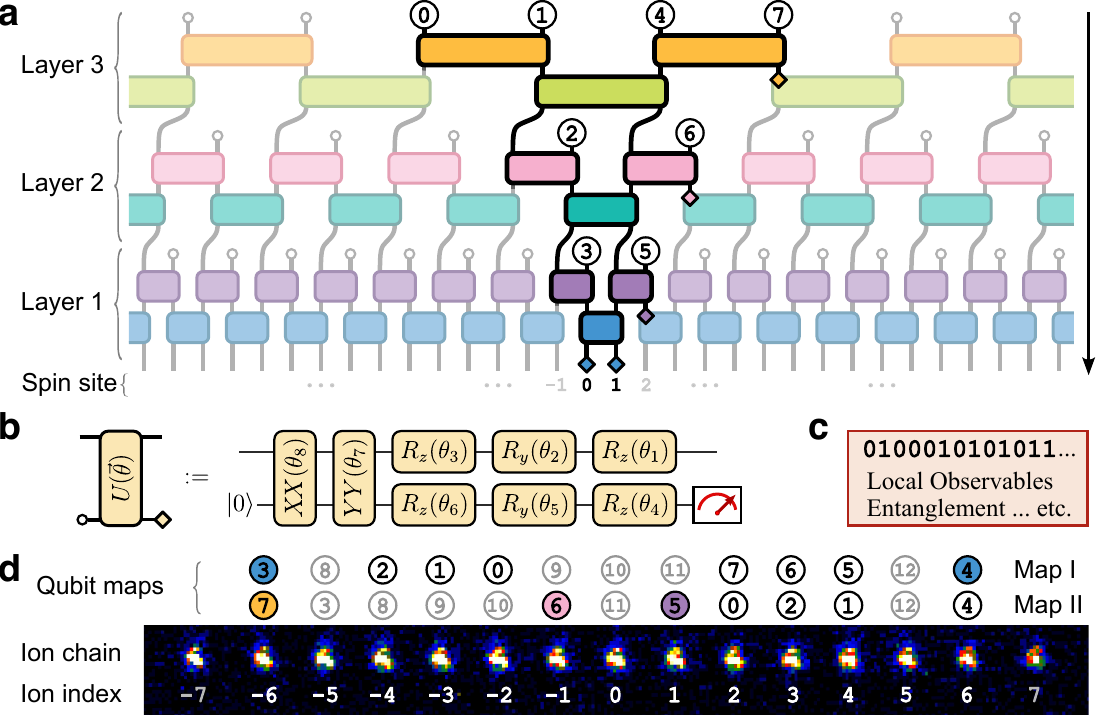}
    \caption{\textbf{MERA on a trapped-ion chain.} \textbf{a} An infinite, binary 1D MERA with three homogeneous layers. The ansatz is prepared from top to bottom as indicated by the arrow, with each layer comprising entangling gates (bare rectangles) and isometries (rectangles with attached circles) that double the number of sites. The output is the many-body state on the physical spin sites. Repeating the MERA circuit horizontally in the spatial dimension generates a state in the thermodynamic limit. Local observables and bipartite entanglement can be studied by executing circuits within the corresponding causal cone (unshaded area) and measuring the target qubits. Circles represent qubits initialized in $|0\rangle$, and diamonds indicate measurements. 
    \textbf{b} Each MERA tensor is implemented as a sequence of two- and one-qubit gates. 
    \textbf{c} Repeated preparation and measurement yield probabilistic binary outcomes (0s and 1s) from which the desired physical properties can be inferred.
    \textbf{d} Fluorescence image of a 15-ion ($^{171}\text{Yb}^{+}$) chain with two hardware-tailored examples of ion-to-qubit mappings: Map I for local measurements on sites 0 and 1, and Map II optimized for entanglement studies with respect to a bipartition into the semi-infinite subsystems $A=(-\infty,1]$ and $B=[2,\infty)$.
    Each qubit is represented by a circle and identified by the number inside. Active qubits in the MERA causal-cone circuit are outlined in black, while unused qubits are shown in gray (shaded). For active qubits, a colored background indicates that the qubit is to be measured, with the color corresponding to the last tensor acting on it.}\label{fig:set-up}
\end{figure}

\section{Setup}\label{sec:setup}
Preparing ground states of many-body systems with long-range quantum fluctuations is challenging. This can be mitigated by accounting for the system's entanglement structure. 
To approximate the ground state $|\Psi\rangle$ in the thermodynamic limit, we implement MERA on a digital ion-trap quantum computer.

\subsection{MERA circuits}

As shown in Fig.~\ref{fig:set-up}a, the layered structure of MERA tensor networks introduces an additional dimension associated with a hierarchy of length and energy scales. The MERA comprises $T$ layers, each composed of unitary (dis)entanglers and isometries, which can be efficiently implemented on our hardware.
Viewed from bottom to top, starting from the physical layer, MERA acts as a renormalization group (RG) flow, where the ascending layer index represents the RG scale.
In each renormalization step, short-distance entanglement is removed by the disentanglers (bare rectangles), and the isometries (rectangles with attached circles) then coarse-grain the system. This process maps physical sites to renormalized effective degrees of freedom, continuing until the final layer $T \sim \log{\xi}$ is reached, where $\xi$ is the largest correlation length in the system. 

Viewed in reverse (the direction of decreasing layer index), the MERA represents a quantum circuit, which prepares the quantum state by first generating long-range correlations associated with low-energy modes, and then progressively incorporating shorter-range correlations that correspond to higher energy scales. For each layer, ancillary qubits initialized in the $|0\rangle$ state (depicted as circles) are injected and entangled with existing qubits through unitary operations. Each unitary tensor consists of two-qubit entangling gates and single-qubit rotations, as illustrated in Fig.~\ref{fig:set-up}b.
\subsection{Observables and holographic subsystem tomography}\label{sec:obs}

Local observables can be studied by implementing only the associated causal-cone circuits. For measurements on sites 0 and 1 of a binary one-dimensional (1D) MERA, the corresponding reduced circuit is indicated by the non-shaded MERA tensors in Fig.~\ref{fig:set-up}a. Due to their isometric properties, the shaded tensors outside the causal cone do not influence the local measurements.

Entanglement entropies and spectra for a spatial bipartition of the system into parts ${A}$ and ${B}$ can be obtained from the reduced density matrix $\hat{\varrho}_{B}=\Tr_{A}|\Psi\rangle\langle\Psi|$. For a MERA, only the gates within the causal cone of the boundary between ${A}$ and ${B}$ need to be implemented, as gates outside the causal cone correspond to unitary basis transformations $\hat{U}_A$ and $\hat{U}_B$ in the two subsystems and do not affect bipartite entanglement properties. Moreover, tomography on the renormalized sites (qubits) at the edges of $A$'s causal cone provides access to $\hat{\varrho}_{B}$ in the basis defined by $\hat{U}_B$.
Thus, due to the causal structure of MERA, the number of gates and number of measured qubits are proportional to the number of layers $T\sim\log\xi$, rather than the total system size and subsystem size, respectively.
Details of our scheme, termed \textit{holographic subsystem tomography}, are described in App.~\ref{sec:HST}.

In the experiment, we implement an infinite, homogeneous, binary 1D MERA with bond dimension $\chi=2$ and up to $T=5$ layers, which captures a maximum correlation range of $3 \times 2^{5}=96$ sites. To this purpose, all tensors of the same color in Fig.~\ref{fig:set-up}a are chosen to be identical, and the shown tensor network structure is continued horizontally ad infinitum.
For the self-similar critical systems, the correlation length $\xi$ diverges, which we address by employing a \textit{scale-invariant MERA}~\cite{Montangero2008-10,Pfeifer2009-79} (see Sec.~\ref{sec:ES} and App.~\ref{sec:scale-invariant}).

\subsection{Trapped-ion system}\label{sec:TI}

We implement MERA circuits on a digital ion-trap quantum computer. Our experimental system consists of a linear chain of fifteen $^{171}\text{Yb}^{+}$ ions, spaced by approximately 3.7\,$\mu$m, confined in a micro-fabricated Paul trap (Sandia HOA-2.1.1~\cite{Maunz2016_01}) held in a room-temperature vacuum chamber. The qubits are encoded in the hyperfine ``clock'' states of the ground $^2\text{S}_{1/2}$ electronic manifold as $|0\rangle \equiv |F=0; m_F=0\rangle$ and $|1\rangle \equiv |F=1; m_F=0\rangle$, where $F$ is the total atomic angular momentum and $m_F$ its projection onto the direction of the magnetic field, with a magnitude of 3.2\,G. Qubits are initialized in the $|0\rangle$ state using optical pumping and measured by state-dependent fluorescence on the 369\,nm D1 line.

The qubits are manipulated via a Raman process using a wide (300\,$\mu$m$\times$30\,$\mu$m) linearly-polarized global-addressing beam and equispaced, linearly-polarized 1\,$\mu$m-diameter individual-addressing beams derived from the same 355-nm pulsed laser (Coherent Paladin Compact 355-4000). Radio-frequency (rf) gate waveforms are produced by an rf system on a chip (ZCU111 from Xilinx Corp.) controlled by the Sandia Octet firmware~\cite{Lobser2020} and separately modulated onto each laser beam using acousto-optic modulators (L3 Harris Corp.). Available single-qubit operations are virtual $z$-axis rotations $R_z(\theta)$, implemented via phase offsets in the radio-frequency controls, and physical $xy$-plane rotations $R(\theta, \phi) = {\rm{e}}^{-{\rm{i}}\theta (\hat{X} \cos{\phi} + \hat{Y} \sin{\phi})/2}$. Entangling gates $X\!X(\theta) = {\rm{e}}^{-{\rm{i}}\theta \hat{X}_i \otimes \hat{X}_j/2}$ between arbitrary qubits $i$ and $j$ are realized through a M\o{}lmer-S\o{}rensen interaction mediated by the shared radial motional modes of the ion chain with mode frequencies near 3 MHz. This interaction is implemented by amplitude-shaping a state-dependent force on the addressed ions using the 355-nm Raman process.

The all-to-all qubit connectivity of our platform simplifies the implementation of the non-local MERA circuits. Using a detailed error model discussed in App.~\ref{sec:system}, 
we determine the ion-to-qubit mapping for each circuit based on the estimated error in the target observables. We suppress crosstalk by assigning the measured qubits to the non-adjacent ions and place higher-fidelity gates closer to the final measurement operations. Based on these principles, in the example of Fig.~\ref{fig:set-up}d, we chose Map~I for measuring local observables and Map~II for measuring entanglement entropy.

\begin{figure*}[t]
    \includegraphics[width=1.8\columnwidth]{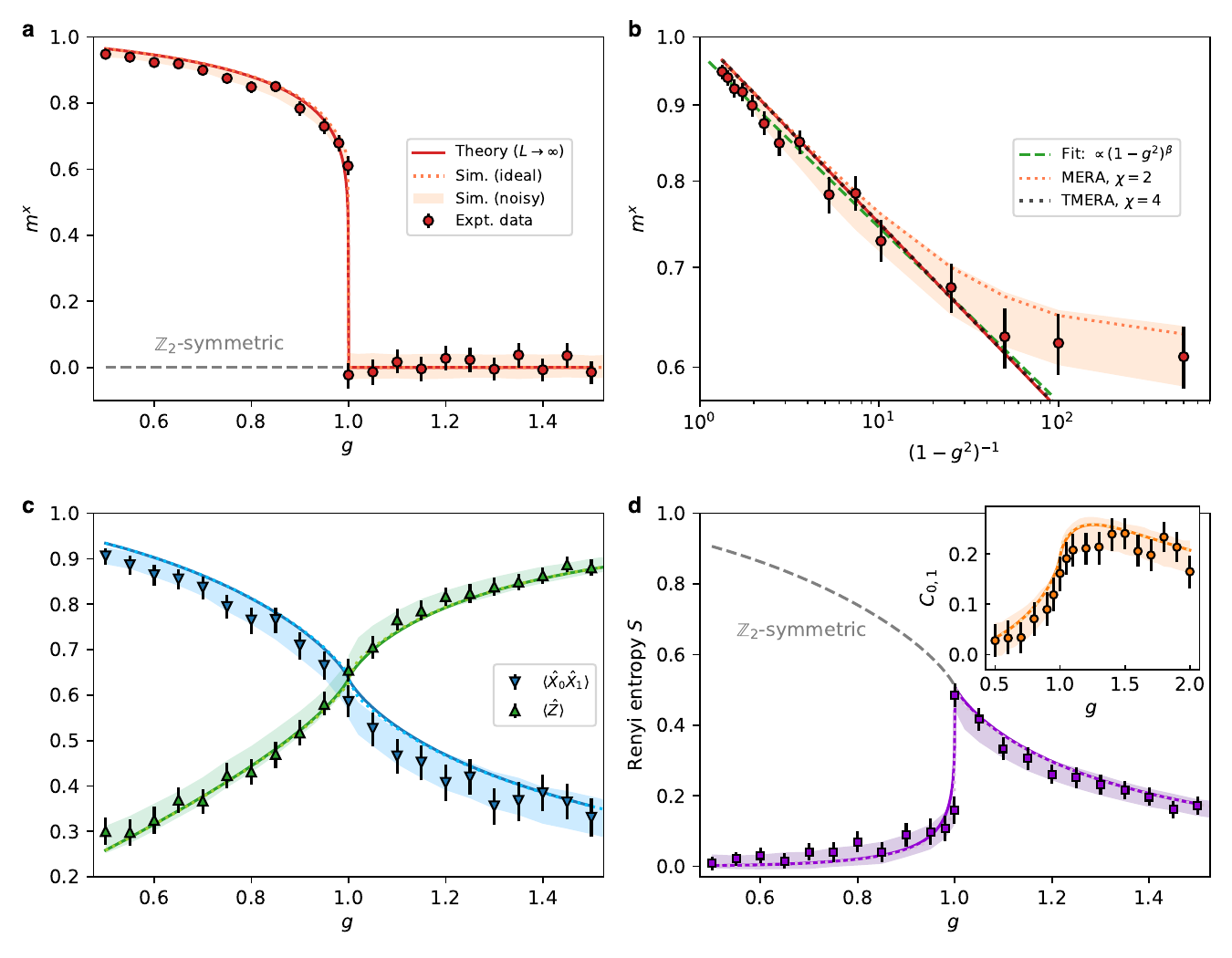}
    \caption{\label{fig:local}\textbf{Local observables and few-site entanglement in the 1D transverse-field Ising model as function of the field $g$.} \textbf{a} Mean magnetization $m^x$ over spin sites 0 and 1. The dashed gray line represents the reference value for a $\mathbb{Z}_2$-symmetric state. \textbf{b} Logarithmic plot of $m^x$ in the $\mathbb{Z}_2$ symmetry-broken ferromagnetic phase ($g<1$). The green dashed line represents a fit to the experimental data, excluding the last two points, yielding an extracted critical exponent of $\beta = 0.117(4)$. For comparison, the exact theoretical value for the Ising class is $1/8=0.125$ (solid red line). The orange and gray dotted lines show the numerical predictions for MERA with bond dimension $\chi=2$ and Trotterized MERA~\cite{Miao2023-5,Miao2023_03} with $\chi=4$, respectively. \textbf{c} Local observables corresponding to the two competing terms in the Hamiltonian \eqref{eq:Ising}. \textbf{d} The R\'enyi-2 entanglement entropy of a single spin. The dashed gray line corresponds to the $\mathbb{Z}_2$-symmetric ground state. The inset shows the nearest-neighbor concurrence $C_{0,1}$ (pairwise entanglement) for two neighboring sites. For all panels, solid lines correspond to theoretical predictions in the thermodynamic limit, dotted lines represent ideal infinite MERA simulations, and the shaded areas indicate the central 95\% quantile obtained from noisy simulations based on an error model that reflects experimental device characteristics. Error bars capture statistical errors, indicating 95\% confidence intervals obtained from a bootstrap resampling of the experimental data, based on 2,000 measurements per quantum circuit.} 
\end{figure*}

\section{Results}\label{sec:results}
To demonstrate the potential of the MERA approach, we apply it to the 1D transverse-field Ising model (TFIM) described by the Hamiltonian
\begin{equation}\label{eq:Ising}
	\hat{H} = -\sum_i^L \hat{X}_i\hat{X}_{i+1} - g\sum_i^L\hat{Z}_i.
\end{equation}
Here, $\hat{X}_i$ and $\hat{Z}_i$ are Pauli operators acting on the spin at lattice site $i$, and $g$ is the transverse magnetic field. The TFIM features a well-characterized continuous quantum phase transition with spontaneous $\mathbb{Z}_2$ symmetry-breaking, making it a good testbed for our approach. The gate compilation of the corresponding MERA tensors and a concrete circuit example are discussed in App.~\ref{sec:circuit}.

Since the 1D binary MERA with small bond dimension is classically simulable, we determine the optimal gate parameters of the variational MERA circuit using classical computers. However, a full hybrid optimization with quantum hardware is also practically feasible (see App.~\ref{sec:MERA-VQE}) and will become necessary when scaling to more complex, higher-dimensional physical models.

\subsection{Quantum phase transition}\label{sec:QPT}
We benchmark the infinite MERA across the quantum phase transition of the TFIM \eqref{eq:Ising} by first measuring local observables $\langle\hat{O}_i\rangle$. We prepare the causal-cone state of the target spins as indicated in Fig.~\ref{fig:set-up}. The number $T$ of employed MERA layers ranges from 2 to 5, depending on the correlation length; $T=2$ layers are used far from criticality and up to 5 layers near the critical point $g=1$. Correspondingly, the number of required qubits ranges from 6 to 12. With these $T$ and bond dimension $\chi = 2$, the numerically optimized MERA already reaches an energy density precision of $\lesssim 10^{-3}$. Thus, unless otherwise noted, the leading errors arise from experimental noise, as we show below.

Figure~\ref{fig:local}a shows the measured magnetization $m^x$ in the $x$-direction as a function of the $z$-field $g$, nicely resolving the phase transition. For $g>1$, the observed magnetization is consistent with zero within the statistical uncertainty, as expected for the paramagnetic phase. For $g<1$, non-zero magnetization is observed along the $x$-direction, indicating spontaneous breaking of the $\hat{X}\to -\hat{X}$ symmetry, characteristic of the ferromagnetic phase. 
In the thermodynamic limit ($L\to\infty$), this phase features two degenerate ground states with opposite spin alignments. Rather than converging to a superposition, the variational optimization of the MERA spontaneously selects one of these symmetry-broken states depending on the random initialization.

Due to experimental noise, the observed magnetization at small $g$ is slightly lower than the theoretical prediction $(1-g^2)^{1/8}$~\cite{McCoy1968-173,Pfeuty1970-57,Osborne2002-66}.
We conducted separate experiments to characterize noise and develop a corresponding error model, as described in App.~\ref{sec:system}. 
In simulations, we model gate-induced dephasing and addressing noise due to ion motion by randomly perturbing gates in each shot. We model idle dephasing, state preparation and measurement (SPAM) errors, and $X$-flip errors during entangling gates by introducing random $Z$- and $X$-flips. The noisy simulations (shaded area in Fig.~\ref{fig:local}a) demonstrate excellent agreement with experimental data.

To determine the critical exponent, we plot the measured order parameter $m^x$ as a function of $(1-g^2)^{-1}$ (Fig.~\ref{fig:local}b). For $(1-g^2)^{-1}<80$, i.e., $g < 0.994$, the experimental data follows the expected power law (solid red line). A fit of the measured $m^x$ in this regime to the form $A\,(1-g^2)^{\beta}$, taking into account the statistical errors, yields the critical exponent of $\beta = 0.117(4)$. This value is close to the theoretical prediction of $1/8=0.125$ for the Ising universality class. As the system enters the immediate vicinity of the critical point ($(1-g^2)^{-1} > 20$), numerical simulations (orange dotted line) show that $m^x$ saturates at a non-zero value. This deviation arises from the small MERA bond dimension $\chi=2$, and can be mitigated by increasing the bond dimension $\chi$. As shown by the thick gray dotted line, a Trotterized MERA circuit~\cite{Miao2023-5,Miao2023_03,Haghshenas2022-12,Haghshenas2024-133} with $\chi = 4$ is predicted to achieve sufficient accuracy to capture the critical behavior even in this regime, at the cost of more qubits and gates.

Figure~\ref{fig:local}c shows the local observables that constitute the Hamiltonian~\eqref{eq:Ising}. As the transverse field $g$ increases, the dominant contribution to the groundstate energy shifts from the correlator $\langle\hat{X}_i\hat{X}_{i+1}\rangle$ to the field term $\langle\hat{Z}_i\rangle$, showing a symmetry due to the Kramers-Wannier duality~\cite{Kramers1941-60} with respect to the critical point $g=1$. The measured values of these observables cross at a slightly smaller value of $g$, aligning with our noise model. Here, the noise is dominated by fluctuations in $X\!X$-gate rotations due to the ions' axial motion
(see App.~\ref{sec:gateError}).

\begin{figure*}[p]
    \includegraphics[width=1.8\columnwidth]{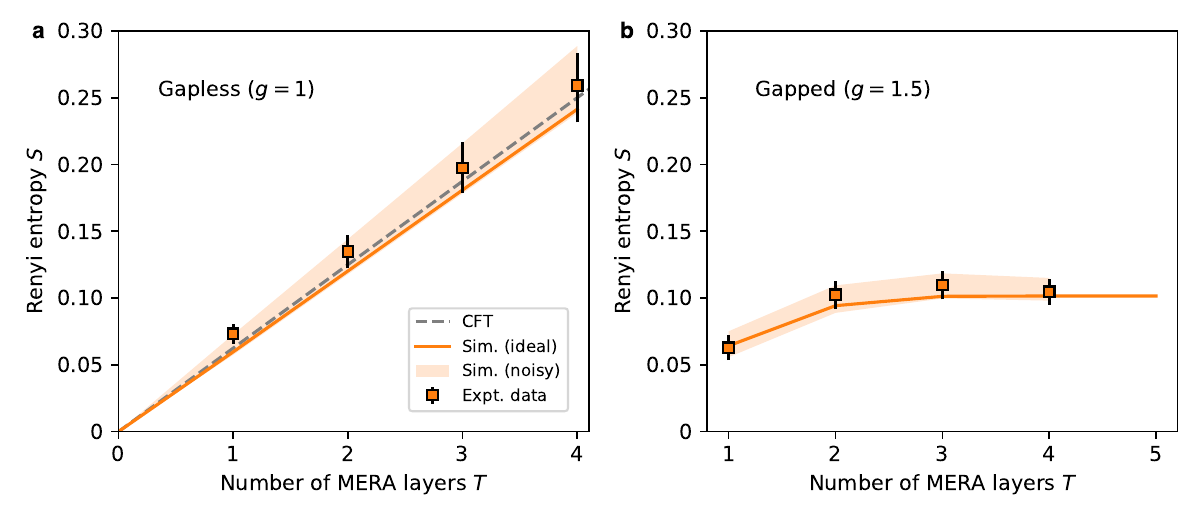}
    \caption{\label{fig:EScaling} \textbf{Log-area-law and area-law scaling of entanglement entropy for critical and gapped systems.} \textbf{a} Bipartite half-chain R\'enyi-2 entanglement entropy for the infinite Ising chain~\eqref{eq:Ising} at the critical point $g = 1$ depending on the number $T$ of MERA layers. The effective subsystem size is $\ell=3\times 2^T$. The gray dashed line shows the log-area law prediction of conformal field theory (CFT) with central charge $c = 1/2$, and the solid line corresponds to ideal MERA simulations. Both the noisy simulation (shaded area) and experimental data (dots with error bars) are shown with 95\% confidence intervals based on 1,500 shots per element of the tomographic measurement basis. \textbf{b} R\'enyi-2 entanglement entropy for MERA states at the non-critical point $g = 1.5$. Here, the half-chain entanglement saturates rapidly as the number $T$ of MERA layers increases, indicating the area law. The shaded area and error bars show 95\% confidence intervals based on 8,000, 5,000, 3,000, and 2,000 shots per measurement basis element for cases with one-, two-, three-, and four-layer MERA, respectively.}
\end{figure*}
\begin{figure*}[p]
    \includegraphics[width=0.95\textwidth]{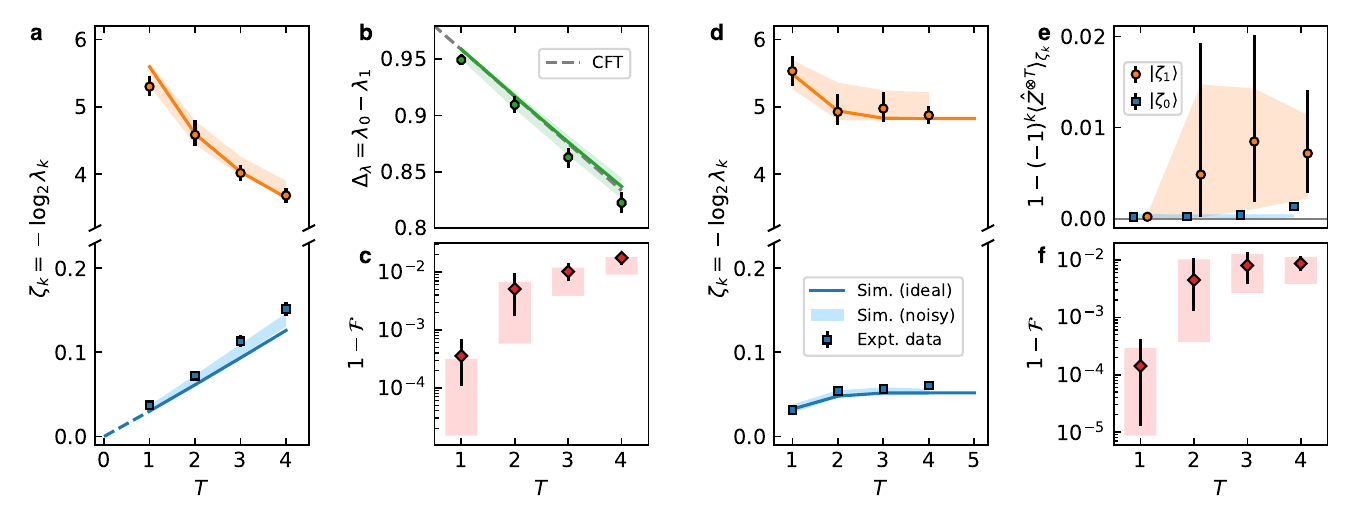}
    \caption{\label{fig:EH} 
    \textbf{Analysis of half-chain density matrices.} (\textbf{a}, \textbf{d}) $T$-dependence of the lowest two eigenvalues $\{\zeta_0,\zeta_1\}$ of the MERA entanglement Hamiltonian at the critical point $g = 1$ and the non-critical point $g=1.5$, respectively. \textbf{b} The Schmidt gap $\Delta_\lambda = \lambda_0 - \lambda_1$ gradually closes as the (sub)system size increases at criticality. The gray dashed line represents the theoretical approximation from Ref.~\cite{DeChiara2012-109}. (\textbf{c, f}) Infidelity of the subsystem density matrices relative to ideal simulations at $g = 1$ and $g=1.5$, respectively. \textbf{e} Expectation values of Pauli $Z$-strings for the first two eigenstates $\{|\zeta_0\rangle,|\zeta_1\rangle\}$ of the entanglement Hamiltonian at $g=1.5$. In all plots, solid lines represent ideal infinite MERA simulations. Both the noisy simulation (shaded area) and the experimental data (dots) are shown with indicators for the 95\% confidence intervals. The results are based on the same dataset as Fig.~\ref{fig:EScaling}, but here, positive semi-definite density matrices are reconstructed using maximum-likelihood estimation.
}
\end{figure*}

\subsection{Entanglement structure}\label{sec:ESt}
In the following, we use MERA to investigate groundstate entanglement across the quantum phase transition in the TFIM~\eqref{eq:Ising}. We find that MERA on the digital quantum computer can resolve details of the entanglement structure, including the hallmark transition from area-law to log-area-law scaling of block entanglement entropies when approaching the critical point.

\subsubsection{Local entanglement measures}\label{sec:Elocal}

For a pure state $|\Psi\rangle$, the entanglement between subsystem ${B}$ and its complement ${A}$ can be quantified by the R\'enyi entanglement entropy
\begin{equation}\label{eq:Renyi}
    S^{(\alpha)}_{B} = \frac{1}{1-\alpha}\log_2 \Tr( \hat{\varrho}_{B}^\alpha),
    \ \text{where}\ 
    \hat{\varrho}_{B} = \Tr_{A} |\Psi\rangle\langle\Psi|   
\end{equation}
is the reduced density matrix of subsystem $B$.
Our approach provides efficient access to the entanglement spectra and, hence, $S^{(\alpha)}_{B}$ for any $\alpha$. 
Here, we focus on the R\'enyi-2 entropy ($\alpha = 2$) as, for the TFIM~\eqref{eq:Ising}, all R\'enyi entropies with $\alpha>0$ exhibit the same scaling.

The reduced density matrix for a single-site is fully characterized by its Bloch vector. The time-reversal symmetry of the TFIM ensures that the local expectation value $\langle \hat{Y}\rangle$ vanishes for all eigenstates. Thus, the Bloch vector is determined by $\langle \hat{X}\rangle$ and $\langle \hat{Z}\rangle$.
Figure~\ref{fig:local}d shows the measured single-site R\'enyi-2 entanglement entropy. Our data faithfully reproduces the predicted sharp peak at criticality and tends to zero for small $g$. In the ferromagnetic phase, the two symmetry-broken ground states are almost product states. Meanwhile, their symmetric superposition is entangled, and the corresponding single-site entanglement is smooth at the critical point.
 
The inset of Fig.~\ref{fig:local}d presents the nearest-neighbor concurrence (pairwise entanglement) $C_{0,1}$~\cite{Osborne2002-66, osterloh2002-416} obtained from the measured nearest-neighbor correlators $\langle\hat{X}_0\hat{X}_1\rangle$, $\langle\hat{Z}_0\hat{Z}_1\rangle$, and $\langle\hat{Y}_0\hat{Y}_1\rangle$~\cite{Syljuasen2003-68, deOliveira2008-77}. The concurrence slightly deviates from theoretical predictions but aligns well with noisy simulations. Notably, the concurrence does not peak at criticality but increases smoothly before reaching a broad maximum in the paramagnetic phase. This suggests that the peak of single-site entanglement at criticality arises not only from short-distance entanglement but also from the accumulation of entanglement over long distances. 
In contrast, in the paramagnetic phase, the nearest-neighbor entanglement is the dominant contribution to the single-site entanglement.

\subsubsection{Entanglement scaling: area-law versus log-area-law}\label{sec:ES}
For the ground states of typical quantum many-body systems in $D$ spatial dimensions, the subsystem entanglement entropy $S(\ell)$ is proportional to the surface area of the subsystem $\propto\ell^{D-1}$, rather than its volume $\propto\ell^D$, where $\ell$ denotes the linear subsystem size~\cite{Eisert2008,Latorre2009,Laflorencie2016-646}. The entanglement entropy is dominated by short-range correlations around the subsystem boundary, resulting in area-law scaling. This can change when the system becomes critical and the correlation length diverges. For critical systems in 1D and fermionic systems with a Fermi surface of codimension $D-1$, the subsystem entanglement entropy is predicted to follow a log-area law $\propto\ell^{D-1}\log{\ell}$~\cite{Srednicki1993,Callan1994-333,Holzhey1994-424,Wolf2005,Gioev2005,Barthel2006-74}.

Resolving these fundamentally different scaling laws has remained an experimental challenge. In small systems, it is difficult to differentiate between log-area-law and area-law scalings. As the linear subsystem size $\ell$ grows, preparing ground states with sufficient precision becomes increasingly difficult, and, in direct approaches, the number of measurements needed for subsystem tomography grows exponentially in $\ell$. 

For MERA, however, each step of the renormalization process changes length scales by a constant factor. This maps a logarithmic scaling in the subsystem size $\ell$ to a linear scaling in the number of renormalization steps $T=1,2,3,\dotsc$. Thus, distinguishing between area and log-area laws in 1D translates to determining whether the entanglement entropy saturates or grows linearly with increasing $T$.

We experimentally realize spatially infinite MERA with a finite number of layers $T$. For a bipartition into subsystems $A=(-\infty, 1]$ and $B=[2,+\infty)$, we implement the holographic subsystem tomography as described
in Sec.~\ref{sec:setup} and detailed in App.~\ref{sec:HST}. 
We prepare the MERA circuit within the causal cone of the subsystem boundary, measure the $T$ renormalized sites at the right edge of the cone (Fig.~\ref{fig:set-up}), and use the classical shadow method~\cite{Huang2020-16} to extract the R\'enyi-2 entanglement entropy \eqref{eq:Renyi}. The finite number of MERA layers limits the maximal correlation length to $3\times 2 ^T$. Thus, we experimentally probe effective subsystem sizes $\ell\propto 2^T$. See
App.~\ref{sec:T-l} 
for a more detailed argument on this relation of $T$ and effective subsystem sizes.

For the non-critical point $g=1.5$ of the TFIM~\eqref{eq:Ising}, the measured half-chain R\'enyi-2 entanglement entropy $S$ is shown in Fig.~\ref{fig:EScaling}b. We observe a rapid saturation of the entanglement entropy for $T \geq 3$, in agreement with the area-law. This is expected as, due to the energy gap, the physical correlation length $\xi$ is finite. Here, additional MERA layers ($T>3$) have a negligible effect on the groundstate approximation and entanglement for this gapped system.
Our error model indicates that, in this regime, SPAM errors and gate rotation errors due to the ions' axial motion are the dominant noise sources.

At criticality ($g=1$), the system has gapless excitations, and the correlation length diverges ($\xi \to \infty$). Here, we model the ground state using a scale-invariant MERA with an infinite number of homogeneous identical layers in both the spatial and preparation directions~\cite{Montangero2008-10,Pfeifer2009-79}. Directly preparing a scale-invariant MERA is not experimentally feasible. Instead, we truncate at layer $T$ and add an additional tensor to the top layer to minimize the finite-$T$ effects as detailed in
App.~\ref{sec:scale-invariant}.

The measured half-chain R\'enyi-2 entanglement entropy $S$ at criticality is shown as a function of $T$ in Fig.~\ref{fig:EScaling}a. The observed entropy increases linearly by approximately $1/16$ per MERA layer. This is consistent with the prediction $\frac{1}{16}\log_2(\ell/a)$ of conformal field theory, where $a$ is an ultraviolet cutoff that regularizes the continuum field theory~\cite{Calabrese2009-42}.
The slight increase in the measured $S$ over the field-theoretical prediction is reproduced by our error model. Here, errors primarily arise during SPAM and from fluctuations in gate angles due to the ions' axial motion; $X$-flips and idle dephasing are secondary
(see App.~\ref{sec:system}).

\subsubsection{Entanglement spectrum}\label{sec:Espec}
Using the holographic tomography data and a maximum-likelihood approach~\cite{Hradil1997-55,Rehacek2001-63,Fiurasek2001-64,James2001-64}, we reconstruct proper (positive semidefinite) density matrices $\hat{\varrho}_B$ 
for subsystem $B$ in the $\hat{U}_B$ basis. They can be written in the form $\hat{\varrho}_B = 2^{-\sum_i{\zeta_i}|\zeta_i\rangle\langle\zeta_i|}$, where $\{\zeta_i := -\log_2{\lambda_i}\}$ is the entanglement spectrum~\cite{Li2008-101} expressed in terms of the eigenvalues $\{\lambda_i\}$ of $\hat{\varrho}_B$.

Figures~\ref{fig:EH}a and~\ref{fig:EH}d show the two lowest entanglement eigenvalues $\{\zeta_0,\zeta_1\}$ of $\hat{\varrho}_B$ for the TFIM \eqref{eq:Ising}, revealing distinct behaviors at and away from the critical point. At criticality ($g=1$), the lowest eigenvalue $\zeta_0$ exhibits a log-law scaling like the entanglement entropy, while $\zeta_1$ decreases with increasing $T$.
As shown in Fig.~\ref{fig:EH}b, the Schmidt gap $\Delta_\lambda = \lambda_0 - \lambda_1$ reduces by approximately 1/24 per renormalization step, which agrees with the approximate theoretical prediction $\Delta_\lambda\sim \ell^{-1/24}$ from Ref.~\cite{DeChiara2012-109}, where $\ell$ denotes again the (effective) subsystem size. For the non-critical point $g = 1.5$, Fig.~\ref{fig:EH}d shows that $\zeta_0$ and $\zeta_1$ saturate with increasing $T$. 

Since $g = 1.5$ lies in the $\mathbb{Z}_2$-symmetric paramagnetic phase, all subsystem density matrices of the ground state must commute with the operator $\otimes_i\hat{Z}_i$. We verify this symmetry by measuring the spin-flip operator on subsystem $B$. As discussed
in Sec.~\ref{sec:setup}, the measurement is done efficiently with the compressed representation of $\hat{\varrho}_B$ on the renormalized sites, exploiting the fact that, for $g\geq 1$, the unitary $\hat{U}_B$ which transforms the measured state to the actual state on $B$ preserves the $\mathbb{Z}_2$-symmetry.
As observed in Fig.~\ref{fig:EH}e, the associated entanglement Hamiltonian exhibits a ground state in the even-parity sector and a first excited state in the odd sector. This mirrors the properties of the system Hamiltonian, aligning with recent efforts concerning connections between entanglement Hamiltonians and system Hamiltonians~\cite{kokail2021-17,Joshi2023-624}. 

\subsubsection{Subsystem fidelity}\label{sec:F}
Finally, we compare the reconstructed subsystem density matrix $\hat{\varrho}_B$ in the $\hat{U}_B$ basis with the ideal MERA simulations, and show the infidelity $1-\mathcal{F}$ in Figs.~\ref{fig:EH}c and~\ref{fig:EH}f for the transverse fields $g = 1.0$ and $1.5$, where the fidelity is defined as
\begin{equation}\label{eq:fd}
    \mathcal{F} = \left(\Tr\sqrt{\sqrt{\hat{\varrho}_B}\,\hat{\rho}_\text{ideal}\,\sqrt{\hat{\varrho}_B}}\right)^2.  
\end{equation}
We achieve infidelities ranging from $10^{-4}$ to $10^{-2}$, which enables the clear discrimination of different entanglement scaling laws, even when the absolute difference in $S$ is small.
These high fidelities arise from careful calibrations, optimized ion-qubit mappings, the intrinsic noise resilience of MERA, suitable gate decompositions for the MERA circuit, and the efficient holographic tomography method. 

Recall that we determine $\hat{\varrho}_B$ through tomography on only $\sim T$ renormalized sites at the edge of subsystem $A$'s causal cone
(cf.~Sec.~\ref{sec:obs} and Fig.~\ref{fig:effMERA}), where, in the basis defined by $\hat{U}_B$, all other qubits are in the $|0\rangle$ state and need not be implemented. This is sufficient because the fidelity metric~\eqref{eq:fd} and bipartite entanglement properties are invariant under the unitary transformation $\hat{U}_B$.

Also, while many widely used variational ansätze employ CNOT gates, our circuits use two-qubit $X\!X$ M{\o}lmer-S{\o}rensen gates with typical absolute rotation angles in the range from $0.01\pi$ to $0.2\pi$, which are much smaller than those of fully entangling gates. As M{\o}lmer-S{\o}rensen gate errors decrease with the rotation angle, this leads to significantly reduced experimental errors.
Appendix~\ref{sec:breakdown} 
provides a breakdown of error contributions and explains how the infidelity scales with experimental errors.

\section{Discussion}
We employed MERA on an ion-trap quantum computer to clearly resolve the quantum phase transition in a condensed matter system, successfully capturing the universal properties of Ising-class physics. Furthermore, our holographic subsystem tomography method made it possible to demonstrate the transition from area-law to log-area-law scaling in groundstate entanglement entropies. This entanglement scaling transition is a hallmark of quantum many-body physics with no classical analog. Its experimental observation had previously faced substantial challenges including finite-size effects and attainable fidelities.

The advantages of the MERA approach become particularly clear when compared to matrix product states (MPS). To capture critical ground states, the sequential geometry of an MPS requires circuit depths scaling as $\mathcal{O}(L)$ and a polynomially growing bond dimension. In contrast, MERA achieves this with a logarithmic depth $\mathcal{O}(\log L)$ and a constant bond dimension owing to its unique hierarchical structure. Finally, while MPS approaches strongly favor 1D systems, MERA captures area-law states across all spatial dimensions $D$ with size-independent bond dimensions~\cite{Barthel2010-105,Evenbly2014-89a}.

The investigation of more complicated models, including systems with higher spin and spatial dimensions, generally requires larger MERA bond dimensions $\chi\gtrsim 4$. As an example, App.~\ref{sec:XXZcircuits} discusses numerical simulations of Trotterized MERA circuits~\cite{Miao2023-5,Miao2023_03} with $\chi= 4$ to demonstrate the Berezinskii-Kosterlitz-Thouless (BKT) phase transition in XXZ spin chains.
In classical MERA simulations, the primary bottleneck are tensor contractions, whose cost scales as $\mathcal{O}(\chi^r)$, where the exponent $r$ depends on the spatial dimension and specific MERA structure. In 1D with $r=7,\dotsc,9$, these costs are still tractable for moderate bond dimensions $\chi\lesssim20$~\cite{Vidal2007-99,Vidal2008-101,Evenbly2009-79}. However, for higher-dimensional systems such as (quasi-)2D frustrated quantum magnets and fermionic systems, where quantum Monte Carlo generally fails due to the negative sign problem, the classical contraction costs with scaling exponents $r=16,\dotsc,26$ become prohibitive even for small $\chi$~\cite{Evenbly2009-79,Barthel2025-111}.

To avoid these high contraction costs, our prior work~\cite{Miao2023_03,Miao2023-5} proposed implementing Trotterized MERA on quantum computers.
As discussed in Refs.~\cite{Miao2023_03,Miao2023-5,Haghshenas2024-133}, the required qubit count grows only logarithmically with $\chi$ and is independent of system size if mid-circuit resets are available. The circuit depth scales as $\mathcal{O}(Tt)$, where the number of MERA layers $T\sim\log\xi$ governs the maximal correlation length $\xi$, and $t$ is a tunable Trotter depth that controls the expressiveness of sub-circuits in individual tensors. For example, simulating a large (e.g., hundreds-by-hundreds) 2D system with a $2\times2\mapsto1$ MERA using $\chi=8$, $T = 6$, and $t = 2$ requires either (i) 108 resettable qubits and a circuit depth of 96 when minimizing depth or (ii) 42 qubits and depth 624 when minimizing the qubit count. Such circuits are not classically tractable, yet they are realistic for state-of-the-art quantum hardware. 

Recent advances in trapped-ion platforms have brought these capabilities closer to realization. For example, 56-qubit systems with $\geq$99.8\% two-qubit gate fidelity are now available~\cite{decross2024_6}, and circuits with over 2000 two-qubit gates have been demonstrated~\cite{haghshenas2025_03}. Larger systems, with $\sim200$ qubits, can be achieved by scaling up the system used in this work~\cite{wu2021tilt}. The primary error source in high-connectivity long ion chains---motional heating of low-frequency modes---can be reduced by over two orders of magnitude using cryogenic techniques and sympathetic cooling with a different isotopic species~\cite{Cetina2022-3}. This approach also facilitates qubit reset and reuse, further enhancing scalability.

Since MERA has been rigorously proven to be free of barren plateaus~\cite{Miao2024-109,Barthel2023_03}, its variational optimization using quantum computers is feasible. More importantly, optimizing MERA with the aid of quantum computers provides a polynomial quantum advantage over classical MERA optimization, which is particularly pronounced in $D\geq 2$ spatial dimensions~\cite{Miao2023_03}.

Once an optimized MERA state is determined, rich physics can be probed through local observables, correlators, and entanglement measures. Measuring local observables and correlators is straightforward, requiring similar resources as energy estimation. Probing entanglement is more demanding, as large-$\chi$ MERA involve multiple qubits per renormalized site, making holographic subsystem tomography costly. To address this, one can perform randomized measurements on edges of the boundary causal cones, and adopt the idea of entanglement Hamiltonian tomography~\cite{Joshi2023-624,kokail2021-17} to learn the renormalized entanglement Hamiltonian, assuming a structured form in terms of renormalized degrees of freedom.

Beyond the study of strongly correlated many-body ground states, MERA states can serve as high-quality starting points for the investigation of nonequilibrium dynamics. As demonstrated in this work, MERA accurately captures both critical and non-critical ground states. A natural extension is to prepare a MERA state and evolve it under quench or Floquet dynamics. Since entanglement growth under nonequilibrium dynamics generally leads to volume-law scaling of entanglement entropies, a quantum advantage is expected as soon as we surpass the realm of exact classical Krylov-subspace methods for $\gtrsim$36 qubits.

\begin{acknowledgments}
We thank K.\ Sun and Y.\ Yu for valuable discussions. 

\end{acknowledgments}

\section*{Funding} 
This work was funded by the NSF Quantum Leap Challenge Institute for Robust Quantum Simulation (OMA-2120757) and the NSF STAQ project (Phy-2325080). Support is also acknowledged from the U.S. Department of Energy, Office of Science, National Quantum Information Science Research Centers, Quantum Systems Accelerator.

\section*{Author contributions} 
Q.M.\ designed the MERA implementation, conducted the numerical simulations, and analyzed the experimental data. T.W.\ executed the experiment and collected the experimental data. T.B.\ initiated the project and devised the holographic subsystem tomography. M.C.\ supervised the experimental work, while T.B.\ and K.R.B.\ supervised the theoretical effort. All authors contributed to the manuscript.

\section*{Competing interests} 
K.R.B. has a personal financial interest in the company IonQ.
M.C. is a co-inventor on U.S. patents US11710061B2, US11152756B2, US11083075B2, and US11573477B2, assigned to University of Maryland. Some of these patents were licensed to IonQ Inc.\ by University of Maryland. Other authors declare that they have no competing interests.

\section*{Data availability} 
All data shown in figures are publicly accessible in the following repository (DOI: \hyperlink{https://doi.org/10.5281/zenodo.20752984}{https://doi.org/10.5281/zenodo.20752984}).
The raw data generated in this study have been deposited in Duke University’s Box cloud storage database. The raw data are available under restricted access subject to intellectual property laws, non-disclosure agreements, and trade-secret protections, in accordance with all applicable laws, regulations, agreement terms and conditions, and the policies and directives of the authors’ sponsors and affiliated institutions. Access can be obtained by contacting Marko Cetina.

\section*{Code availability}
The MERA code used in this work is publicly accessible in the following repository (DOI: \hyperlink{https://doi.org/10.5281/zenodo.20690576}{https://doi.org/10.5281/zenodo.20690576}).

\begin{figure*}[p]
    \includegraphics[width=0.9\textwidth]{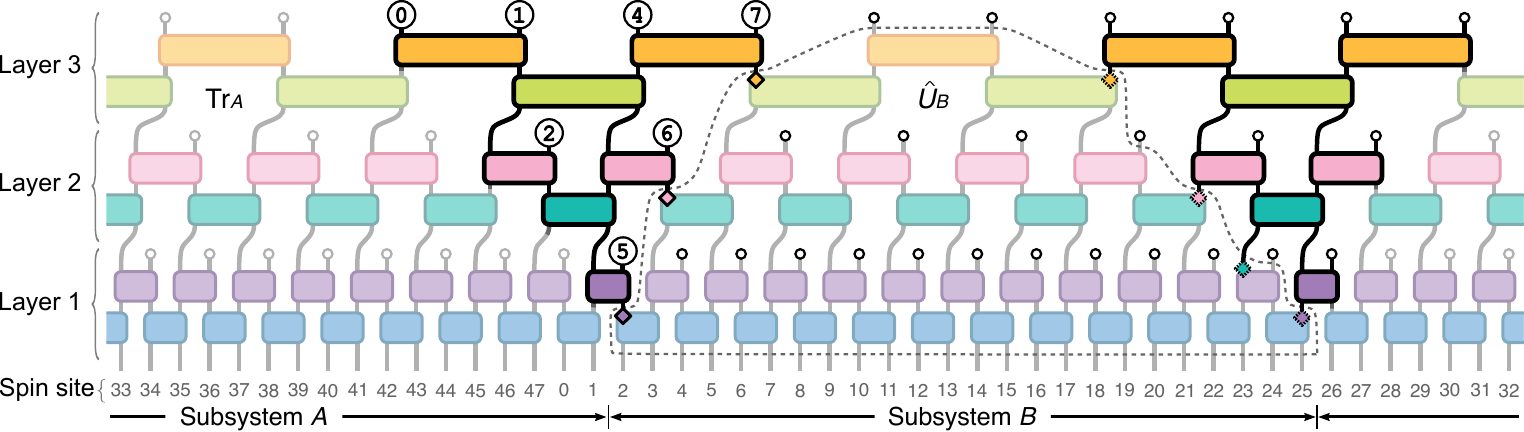}
    \caption{\label{fig:effMERA}\textbf{Holographic subsystem tomography and entanglement scaling.} An infinite binary MERA with $T$ layers is consistent with a MERA on $6\times2^T$ sites with periodic boundary conditions. Here, $T=3$, resulting in a total of $6\times 2^3=48$ sites. All connected two-point correlation functions for sites of distance $\geq 3\times2^T-2=22$ are exactly zero.
    For example, $\langle\hat{X}_{10}\hat{Y}_{31}\rangle-\langle\hat{X}_{10}\rangle\langle\hat{Y}_{31}\rangle$ can be nonzero while $\langle\hat{X}_{10}\hat{Y}_{32}\rangle-\langle\hat{X}_{10}\rangle\langle\hat{Y}_{32}\rangle\equiv 0$. This is due to the fact that, in the latter case, the causal cones of the two operators have no overlap. Similarly, for any bipartition of the system into equally sized blocks like $A=\{26,\dotsc,47,0,1\}$ and $B=\{2,\dotsc,25\}$, the causal cones of the two subsystem boundaries do not overlap. Hence, measurements of the renormalized sites at the two boudnaries are independent of each other, i.e., follow a product distribution.}
\end{figure*}
\begin{figure*}[p]
    \includegraphics[width=0.9\textwidth]{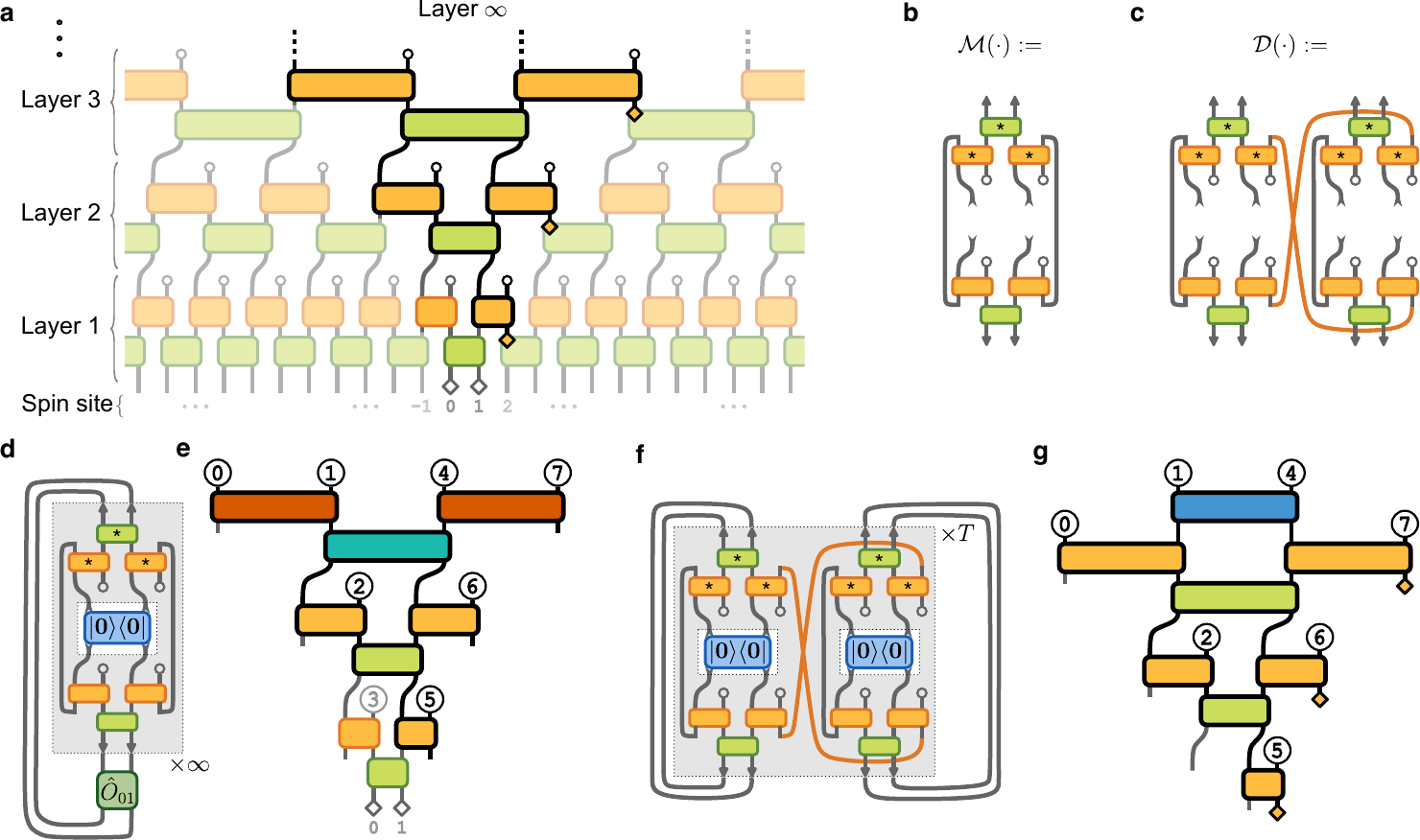}
    \caption{\label{fig:SI-mera} \textbf{Scale-invariant MERA and its implementation on the quantum computer.} \textbf{a} A scale-invariant binary 1D MERA with an infinite number of lattice sites and layers. All unitaries (bare rectangles) and isometries (rectangles with attached circles) are identical across layers and the spatial direction. \textbf{b} Local observables on sites 0 and 1 have a causal cone consisting of a repeating configuration of tensors in each layer. The corresponding repeated layer-transition channel $\mathcal{M}(\cdot)$ maps reduced density matrices from the top to the bottom layer. Stars indicate complex conjugation. \textbf{c} The shown doubled and SWAP-contracted layer-transition map $\mathcal{D}$ can be used to compute the second-order R\'enyi entanglement entropy of a bipartition into subsystems $(-\infty,1]$ and $[2,\infty)$ (see panel \textbf{f}). \textbf{d} Causal-cone tensor-network for the evaluation of local expectation values on sites 0 and 1. The MERA circuit is initialized with reference state $|0\rangle$ at the top layer. \textbf{e} In the experiment, we truncate the scale-invariant MERA after layer $T$. For the measurement of local observables, the top layer is optimized to approximate the dominant eigenmode $\hat{\rho}_\text{steady}:=\lim_{T\to\infty}\mathcal{M}^{T}(|\bf{0}\rangle\langle \bf{0}|)$ of the layer-transition channel. Finally, we measure the qubits at sites 0 and 1. \textbf{f} The tensor network for the purity of the reduced density operator of subsystem $[2,\infty)$. \textbf{g} In this case, the top tensor (blue) is optimized to suppress sub-leading contributions from the eigenvectors of the doubled map $\mathcal{D}$. Scale-invariant layers are then appended. To extract the scaling of entanglement as a function of the number $T$ of RG steps, we perform tomography on $T$ qubits at the right edge of the boundary causal-cone circuit (see also Fig.~\ref{fig:effMERA}).}
\end{figure*}

\appendix

\section{Holographic subsystem tomography and entanglement scaling}\label{sec:HST}
For the study of bipartite entanglement entropies, entanglement spectra, and subsystem state fidelities as discussed in 
Sec.~\ref{sec:results}, we consider MERA approximations $|\Psi\rangle$ for the ground state and a spatial bipartition into subsystem $A$ and its complement $B$.

The goal of the holographic subsystem tomography is to efficiently access the reduced density matrix $\hat{\varrho}_B = \Tr_A|\Psi\rangle\langle\Psi|$ of subsystem $B$ in the experiment. The isometric property ($\hat{W}^\dag\hat{W}=\mathbbm{1}$) of the MERA tensors implies that, under the partial trace $\Tr_A$, all tensors outside the causal cone of $B$ can be removed, as exemplified by the shaded area atop subsystem $A$ in Fig.~\ref{fig:effMERA}.
Furthermore, the aforementioned quantities are all invariant under unitary transformations on subsystem $B$. Using this invariance, we can remove all MERA gates outside the causal cone of subsystem $A$, which corresponds to the action of the unitary transformation $\hat{U}_B^\dag$ from Sec.~\ref{sec:setup}. 
Only the gates inside the causal cone(s) of the subsystem boundary(ies) remain. All qubits outside this causal cone are in the reference state $|0\rangle$ and need not be implemented in the experiment. 

To determine the transformed subsystem density matrix
\begin{equation}\label{eq:rhoB-eff}
    \hat{U}_B^\dag \hat{\varrho}_B \hat{U}_B = \Tr_A( \hat{U}_B^\dag |\Psi\rangle\langle\Psi| \hat{U}_B),
\end{equation}
we only need to measure the $\sim T$ non-trivial renormalized sites at the edge of $A$'s causal cone, as indicated by the diamonds in Fig.~\ref{fig:effMERA}. Again $T$ denotes the number of MERA layers. 

To further reduce experimental resource requirements, we consider either infinite chains with a bipartition into two semi-infinite subsystems, or finite chains with periodic boundary conditions where block $B$ is large enough such that the causal cones of its two boundaries are disjoint as exemplified in Fig.~\ref{fig:effMERA}. In the latter scenario, the transformed density operator \eqref{eq:rhoB-eff} is actually a tensor product of a state $\tilde{\varrho}_B^L$ at the left edge, a state $\tilde{\varrho}_B^R$ at the right edge, and central sites in the reference state $|0\rangle\langle 0|$,
\begin{equation}\label{eq:rhoB-PBC}
    \hat{U}_B^\dag \hat{\varrho}_B \hat{U}_B = \tilde{\varrho}_B^L\otimes|0\rangle\langle 0|\otimes\dotsc\otimes|0\rangle\langle 0|\otimes \tilde{\varrho}_B^R.
\end{equation}
Due to the product structure, $\tilde{\varrho}_B^L$ and $\tilde{\varrho}_B^R$ can be measured independently.

Moreover, when studying quantities that only depend on the spectrum of $\hat{\varrho}_B$
for a bipartition of the system into two equally sized halves and for total system sizes $L=6\times 2^T,8\times 2^T,...$, it is sufficient to measure only one of the two density operators. In this case, the MERA circuits for $\tilde{\varrho}_B^L$ and $\tilde{\varrho}_B^R$ prepare the same pure state, and only differ in the selection of sites to be measured: those at the right edge of the causal cone for $\tilde{\varrho}_B^L$ and those at the left edge for $\tilde{\varrho}_B^R$, as indicated in Fig.~\ref{fig:effMERA}. Thus, all nonzero eigenvalues of $\hat{\varrho}_B$ are given by the elementwise product
\begin{equation}\label{eq:specProduct}
    \{\lambda_i\lambda_j\}\quad\text{of}\ \tilde{\varrho}_B^L\text{'s spectrum} \ \ \{\lambda_i\}    
\end{equation}
with itself.

\section{Subsystem size versus number of MERA layers}\label{sec:T-l}
In the main text, we associated the number $T$ of layers in infinite MERA with the logarithm of subsystem sizes $\ell$. Here, we elaborate further on this relation.

An infinite binary 1D MERA $|\Psi\rangle$ with $T$ layers have a maximum correlation range $\xi_\text{max} := 3\times2^T$, where the prefactor 3 arises from the maximal causal-cone width of local operators acting on $k\leq3$ contiguous sites. In particular, connected correlation functions
\begin{equation}
    \langle\hat{O}_i\hat{O}'_{i+\delta\ell}\rangle_\Psi - \langle\hat{O}_i\rangle_\Psi\langle\hat{O}'_{i+\delta\ell}\rangle_\Psi   
\end{equation}
are exactly zero for all $i$ and local operators $\hat{O}_i$ and $\hat{O}'_j$ whenever $\delta\ell\geq \xi_\text{max}-3+k$.

Hence, all local observables, two-point correlation functions, and subsystem density matrices for blocks of $\ell\leq \xi_\text{max}$ sites of an infinite MERA coincide with those of a corresponding finite-size MERA for a system of $2\xi_\text{max}=6\times2^T$ sites with periodic boundary conditions.

Furthermore, the half-chain density operator of an infinite $T$-layer MERA, considered in Sec.~\ref{sec:results}, is given by
\begin{equation}\label{eq:rhoB-infinte}
    \hat{U}_B^\dag \hat{\varrho}_B \hat{U}_B = \tilde{\varrho}_B^L\otimes|0\rangle\langle 0|\otimes|0\rangle\langle 0|\otimes\dotsc,
\end{equation}
where $\tilde{\varrho}_B^L$ is the same state as the one in Eq.~\eqref{eq:rhoB-PBC} for an $\ell=3\times2^T$-site block in a periodic $6\times2^T$-site system, establishing the exponential relation between $\ell$ and $T$. The spectra of the states \eqref{eq:rhoB-infinte} and \eqref{eq:rhoB-PBC} are related by the elementwise product \eqref{eq:specProduct}.

\section{Scale-invariant MERA}\label{sec:scale-invariant}
Scale-invariant MERA are designed for critical systems, where correlation lengths diverge. As shown in Fig.~\ref{fig:SI-mera}a, they feature an elementary cell of tensors that repeats in the spatial and renormalization directions, in accordance with the self-similarity of critical systems.

Viewed in the renormalization direction, the scale-invariant MERA maps the local Hamiltonian of a critical system into an RG fixed point. Seen in reverse, the scale-invariant MERA acts as a quantum channel that iteratively refines the state, starting from an initial reference state and adding layers to construct the ground state of the critical system.

Consider spin sites 0 and 1. Their causal-cone state as indicated by the unshaded region in Fig.~\ref{fig:SI-mera}a
is constructed by applying the same layer-transition channel $\mathcal{M}$ (Fig.~\ref{fig:SI-mera}b) multiple times, where renormalized sites that leave the causal cone are traced out. $\mathcal{M}$ is a quantum channel and can be diagonalized in bi-orthonormal operator bases such that
\begin{equation}\label{eq:M}
    \mathcal{M}(\cdot) = |\hat{\varrho}_\text{steady}\rangle\!\rangle\langle\!\langle\mathbbm{1}| + \sum_{i>0}m_i |\hat{r}_i\rangle\!\rangle\langle\!\langle \hat{\ell}_i|,    
\end{equation}
where $|m_i|<1$ for all $i>0$, and we use a super-bra-ket notation for the operator-basis elements according to the Hilbert-Schmidt inner product $\langle\!\langle\hat{L}|\hat{R}\rangle\!\rangle:=\Tr(\hat{L}^\dag\hat{R})$. With $\hat{r}_0\equiv \hat{\varrho}_\text{steady}$ and $\hat{\ell}_0\equiv \mathbbm{1}$, $\hat{r}_i$ and $\hat{\ell}_i$ are the right and left eigen-operators of $\mathcal{M}$ for eigenvalue $m_i$ and obey $\langle\!\langle\hat{\ell}_i|\hat{r}_j\rangle\!\rangle=\delta_{i,j}$.

We initialize the MERA in the product state $|\mathbf{0}\rangle=|0,\dotsc,0\rangle$ at the topmost (infinite) layer. The expectation value of a local observable $\hat{O}_{0,1}$ with the scale-invariant MERA then is
\begin{equation}
  \lim_{T\to\infty}\Tr\left(\hat{O}_{0,1}\mathcal{M}^T(|\bf{0}\rangle\langle\bf{0}|)\right) = \Tr(\hat{O}_{0,1}\hat{\varrho}_\text{steady}).
\end{equation}
See Fig.~\ref{fig:SI-mera}d. Experimentally, we cannot directly implement MERA with an infinite number of layers. Instead, we truncate after layer $T$ and use the top layer to prepare a two-site state $\hat{\varrho}^\prime$ that maximizes the contribution of the dominant eigenmode $\hat{\varrho}_\text{steady}$ of the layer-transition channel \eqref{eq:M}. This is achieved by choosing the gate angles $\{\bf{\theta}\}$ in layer $T$ as $\arg\min_{\bf{\theta}} \sum_{i>0}|m_i\langle\!\langle\hat{\ell}_i|\hat{\varrho}^\prime\rangle\!\rangle|^2$. We then append $T-1$ layers of the scale-invariant MERA to refine the dominant eigenmode as shown in Fig.~\ref{fig:SI-mera}e. This approach is highly effective as demonstrated for $T=5$ and a few local observables in Secs.~\ref{sec:QPT} and \ref{sec:Elocal}.

To account for layer truncations in the study of subsystem purity and entanglement, we introduce a doubled SWAP-contracted layer-transition map $\mathcal{D}$ as shown in Fig.~\ref{fig:SI-mera}c. For a scale-invariant MERA truncated after layer $T$ and initialized with the reference state $|\bf{0}\rangle$, the purity of subsystem $B=[2,\infty)$ can be computed by iteratively applying $\mathcal{D}$ to the (doubled) zero-product reference state $|\bf{0},\bf{0}\rangle\langle\bf{0},\bf{0}|$ at the topmost layer and taking the final trace as
\begin{equation}
  \Tr(\hat{\varrho}_B^2)=\Tr\left(\mathcal{D}^T(|\bf{0},\bf{0}\rangle\langle\bf{0},\bf{0}|)\right);
\end{equation}
see Fig.~\ref{fig:SI-mera}f. Similarly to the channel $\mathcal{M}$, we can diagonalize $\mathcal{D}$
such that $\mathcal{D}(\cdot) = \sum_i d_i|\hat{R}_i\rangle\!\rangle\langle\!\langle \hat{L}_i|$ with $|d_0|>|d_i|$ for all $i>1$. Due to the SWAP operation, $\mathcal{D}$ is not a quantum channel and $|d_0|<1$. The second-order R\'enyi entanglement entropy with $\alpha=2$ in Eq.~\eqref{eq:Renyi} evaluates to 
\begin{eqnarray}\nonumber
    S^{(2)}_{B} &=& -\log_2 \left( \Tr\left(\mathcal{D}^T(|\bf{0},\bf{0}\rangle\langle\bf{0},\bf{0}|)\right) \right) \\\nonumber
    &=& - \log_2\left( \sum_{i=0} d_i^T \langle\!\langle\mathbbm{1}|\hat{R}_i\rangle\!\rangle \langle\!\langle \hat{L}_i|\bf{0},\bf{0}\rangle\!\rangle \right) \\
    &\to & -T\log_2{d_0}-\log_2\left(\langle\!\langle\mathbbm{1}|\hat{R}_0\rangle\!\rangle \langle\!\langle \hat{L}_0|\bf{0},\bf{0}\rangle\!\rangle\right),\label{eq:Strunc}
\end{eqnarray}
where the large-$T$ limit is taken in the last line, and $|\bf{0},\bf{0}\rangle\!\rangle$ represents the projector $|\bf{0},\bf{0}\rangle\langle\bf{0},\bf{0}|$. Hence, the entanglement entropy should increase linearly with $T$, i.e., by a constant in each renormalization step.

Even with the small bond dimension $\chi=2$, the dominant eigenvalue $d_0$ of $\mathcal{D}$ in the optimized scale invariant MERA agrees well with the conformal field theory prediction for the Ising model with central charge $c=1/2$:
\begin{equation}
   -8\log_2d_0 = 0.475\ldots \approx c .
\end{equation}

\begin{figure*}
    \centering
    \includegraphics[width=0.95\textwidth]{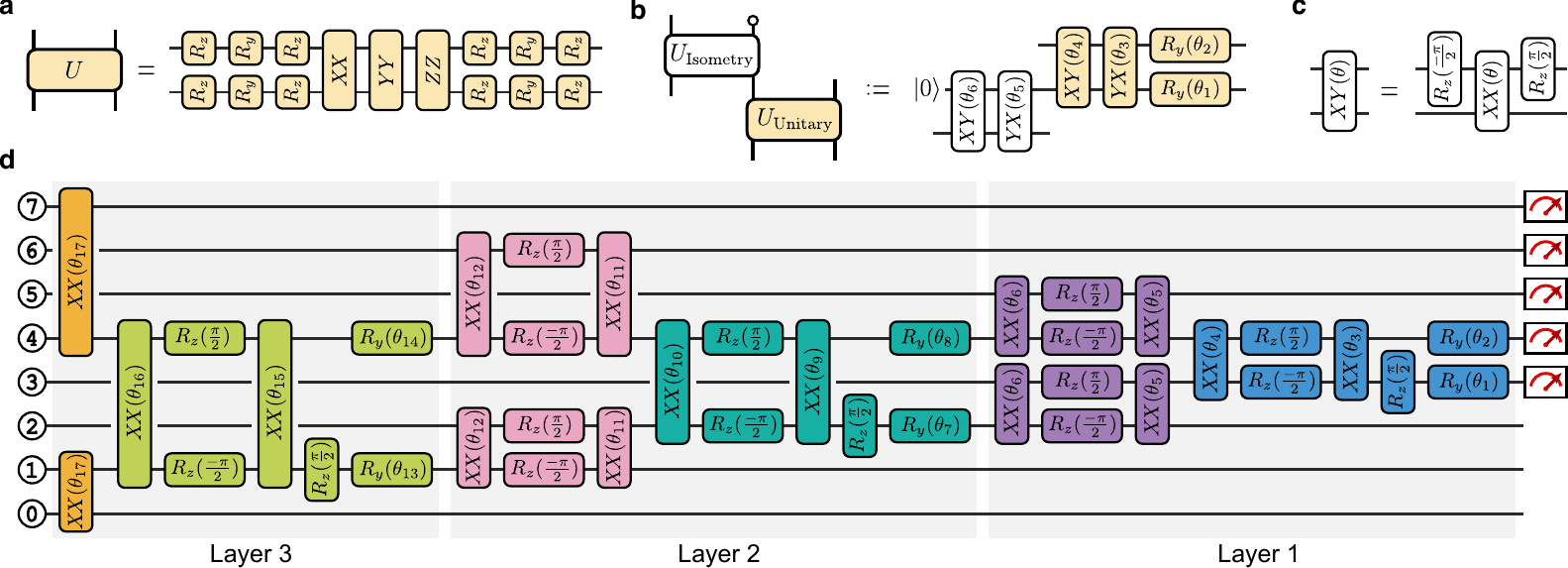}
    \caption{\textbf{Gate compilation and circuit examples.} \textbf{a} Gate decomposition of a generic two-qubit unitary. \textbf{b} Gate compilation for a unit cell in the binary 1D MERA, incorporating symmetry constraints. The isometry tensor is formed by injecting a $|0\rangle$ qubit (indicated by a circle) into one leg of a unitary tensor. The unit cell contains six variable gates. \textbf{c} Equivalence between the $XY$ gate and a circuit of three native gates. \textbf{d} An example circuit used in the experiment, composed of native gates. Qubits (circles with indices) are initialized in the $|0\rangle$ state. The circuit consists of three layers, and in each layer, gates of the same color correspond to either two isometric tensors or a single unitary (dis)entangler, respectively. Local observables at spin sites 0 and 1 are obtained by measuring qubits 3 and 4, while subsystem entanglement is extracted by measuring qubits 5, 6, and 7.}
    \label{fig:circuits}
\end{figure*}

Again, we truncate the scale-invariant MERA after layer $T\leq 4$ for experiments. To mitigate the effect of the truncation, we do not initialize with the reference state $|\bf{0}\rangle$ at the top layer but with a more suitable two-site state $|\varphi\rangle$, generated from $|\bf{0}\rangle$ through one entangling and further single-qubit gates. These gates are chosen to suppress subleading contributions in \eqref{eq:Strunc} by minimizing $\sum_{i>0}|d_i\langle\!\langle\mathbbm{1}|\hat{R}_i\rangle\!\rangle\langle\!\langle \hat{L}_i|\varphi,\varphi\rangle\!\rangle|^2$. Thus, we arrive at the approximation 
\begin{equation}
    S^{(2)}_{B} = -\log_2(\Tr(\mathcal{D}^T(|\varphi,\varphi\rangle\langle\varphi,\varphi|)))\approx - T\log_2d_0.
\end{equation}

After implementing such $T$-layer approximations of a scale-invariant MERA in the experiment, we perform the holographic subsystem tomography, measuring all $T$ qubits at the right edge of the boundary causal-cone as discussed in 
Sec.~\ref{sec:HST} 
and illustrated in Fig.~\ref{fig:SI-mera}g.

\section{Gate Compilation and Circuit Examples}\label{sec:circuit}
In the main text, we showed tensor network diagrams of MERA and schematic gate decompositions of individual tensors. Here, we demonstrate how to compile these into native quantum gates and provide concrete circuit examples for measuring both local observables and nonlocal entanglement properties.

As shown in Fig.~\ref{fig:circuits}a, a generic 4-by-4 unitary operation can be decomposed into a two-qubit circuit using 3 entangling gates and 12 single-qubit gates, involving a total of 15 parameters (up to a global phase). However, this generic decomposition turns out to be unnecessary for our MERA circuits. For example, in a network of such two-qubit unitaries, redundant single-qubit gates on connected bonds can be eliminated. More importantly, a physically motivated ansatz should preserve the symmetries of the Hamiltonian, except in cases of spontaneous symmetry breaking.

For the transverse-field Ising model studied here, the ground state exhibits time-reversal symmetry (invariance under complex conjugation in the computational basis) for all values of $g$ and a $\mathbb{Z}_2$ symmetry when $g\leq1$. While a global symmetry does not necessarily imply local symmetry, we find that, in our case, it does hold locally. A suitable choice of entangling gates that respect the symmetries of the Ising model includes $XY(\theta) = {\rm{e}}^{-{\rm{i}}\theta \hat{X} \otimes \hat{Y}/2}$ and $Y\!X(\theta) = {\rm{e}}^{-{\rm{i}}\theta \hat{Y} \otimes \hat{X}/2}$, which are real in the computational basis
and respect the model's $\mathbb{Z}_2$ symmetry as $[\otimes_i \hat{Z}_i, XY(\theta)] = 0$. Based on this, we replace $X\!X$ and $YY$ gates with $XY$ and $Y\!X$ gates. We remove $ZZ$ and single-qubit $Z$ gates as they violate time-reversal symmetry. Variable $R_y(\theta)={\rm{e}}^{-{\rm{i}}\theta \hat{Y}/2}$ gates are retained in some tensors to allow for spontaneous breaking of the $\mathbb{Z}_2$ symmetry. The $R_y$ gates that have been removed only have a relatively small effect on the accuracy.

\begin{figure*}
    \centering
    \includegraphics[width=0.9\textwidth]{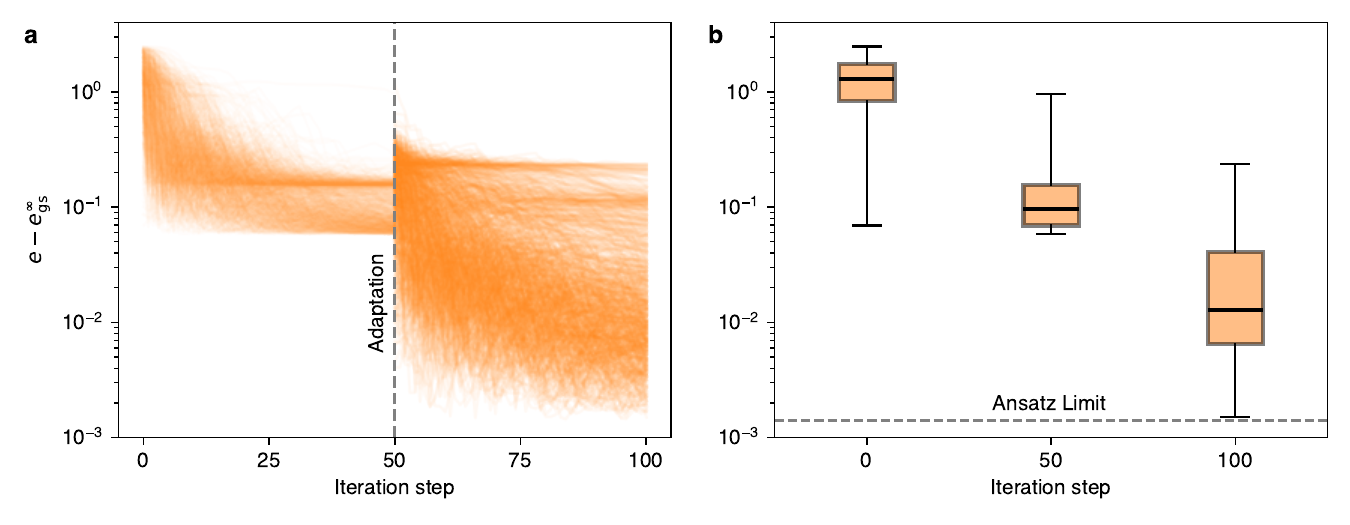}
    \caption{\label{fig:vqe} \textbf{Simulation of hybrid variational optimization.} \textbf{a} Energy‑density error $e - e^\infty_{\mathrm{gs}}$ depending on the iteration step for 1,000 randomly initialized MERA circuits. \textbf{b} Quantile distributions at different optimization steps. The whisker end caps indicate the minimum and maximum values, the box edges denote the 25\% and 75\% quantiles, and the black line marks the median. The gray dashed line indicates the ansatz limit after adaptation (second stage).} 
\end{figure*}

The resulting refined gate compilation for a unit cell in the binary MERA, including one isometry and one unitary (dis)entangler, is shown in Fig.~\ref{fig:circuits}b.
This symmetry-respecting compilation significantly reduces the parameter space and redundancy without sacrificing expressiveness in approximating the ground state, as confirmed by our numerical simulations. Since binary MERA with one qubit per bond is classically simulable, we do not implement the variational MERA on the quantum computer. Instead, we classically optimize the MERA with the gate configuration in Fig.~\ref{fig:circuits}b. In the classical optimization procedure, convergence is reached when the energy density no longer decreases and fluctuates only within a small tolerance (typically on the order of $10^{-8}$ or below). We find that the optimization process is robust and yields consistent results across different initial conditions. Examining these optimized parameters, we note that while the $R_y$ gate is retained for all values of $g$, the optimized angles are non-zero only when $g < 1$ within our data points, which is consistent with the presence of a non-zero order parameter in the ferromagnetic phase.

In our experimental setup, the native entangling gate is the $X\!X$ gate. As shown in Fig.~\ref{fig:circuits}c, we convert the $XY$ and $Y\!X$ gates into native gates by appending additional virtual $R_z(\pm\frac{\pi}{2})$ rotations, which are effectively free, as mentioned in
Sec.~\ref{sec:TI} of 
the main text.

As previously discussed and shown in Fig.~\ref{fig:set-up}, local observables and subsystem entanglement can be probed by preparing only the causal cone state (unshaded regions in Figs.~\ref{fig:set-up} and \ref{fig:effMERA}), rather than the full MERA circuit. Figure~\ref{fig:circuits}d presents a concrete example of such a circuit. From left to right, 8 qubits are initialized in the $|0\rangle$ state, followed by layers of gates grouped into either isometry or (dis)entangler tensors. Finally, the desired observables are measured. In this circuit example, measurements on qubits 3 and 4 yield local observables on spin sites 0 and 1 as indicated in Fig.~\ref{fig:set-up}, while qubits 5, 6, and 7 are used to evaluate the entanglement between the subsystems $A$ and $B$ consisting of spin sites $(-\infty,1]$ and $[2,\infty)$, respectively.

\section{Variational optimization of MERA circuits}\label{sec:MERA-VQE}

As discussed in the main text, since the 1D MERA circuits utilized in our experiment are classically simulable, we optimized them via classical computation. However, the classical numerical cost of tensor contractions increases sharply for dimensions $D > 1$. By replacing these classical computations with quantum circuits on a quantum processor, a hybrid optimization of MERA circuits offers a scalable pathway for the investigation of many-body ground states. To demonstrate the feasibility of this framework and explore its behavior under realistic experimental conditions, we present a corresponding numerical simulation of the hybrid variational optimization.

As a concrete example, we consider the critical point $g = 1.0$ of the 1D Ising model and apply a scale‑invariant MERA ansatz. The true ground state at criticality is strongly correlated and challenging to approximate. Nevertheless, a scale‑invariant MERA with the smallest bond dimension $\chi = 2$ can already accurately represent this state, achieving an energy‑density error as low as $1.4 \times 10^{-3}$ in noiseless numerical simulations.

To reflect experimental conditions, our variational optimization simulations explicitly incorporate measurement uncertainty by sampling measurement outcomes rather than using exact energies. To reduce the sampling complexity, we employ the SPSA (Simultaneous Perturbation Stochastic Approximation) optimizer instead of standard gradient descent, which requires evaluating the energy at only two parameter points per iteration.

\begin{figure*}
    \centering
    \includegraphics[width=0.9\textwidth]{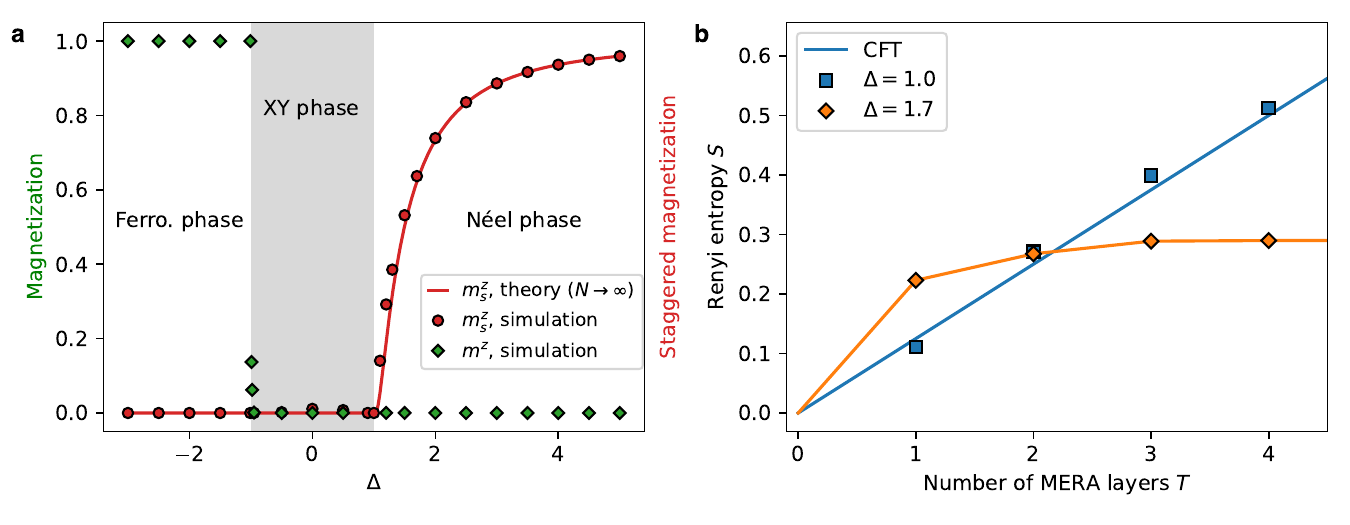}
    \caption{\label{fig:XXZ} \textbf{XXZ chain simulated via quantum MERA circuits.} (a) The staggered magnetization $m^z_s$ serves as an order parameter for the $\mathbb{Z}_2$ symmetry breaking when $\Delta > 1$. The uniform magnetization $m^z$ equals 1 in the ferromagnetic phase and 0 otherwise. (b) Entanglement scaling at and away from criticality. The solid line represents the conformal field theory (CFT) prediction across the entire XY phase, while the dots show the scaling obtained from MERA circuits.} 
\end{figure*}

Figure~\ref{fig:vqe}a shows the energy‑density error during optimization for 1,000 randomly initialized MERA circuits. To improve convergence and reduce resource costs, we adopt an adaptive strategy. During the first 50 iterations, we optimize a single-layer MERA ($T = 1$) using only 100 shots per measurement basis. Since the Ising Hamiltonian only involves the $Z$ and $X$ bases, each iteration requires 400 shots in total to evaluate the gradient and determine the move direction. In this stage, the energy error typically reaches the 0.1--0.2 range, consistent with the shot‑noise limit. In the subsequent 50 iterations, we increase the number of MERA layers to $T = 6$ and the shot count to 10,000 per measurement basis. By the end of this stage, most trajectories converge to energy errors on the order of $10^{-2}$, again consistent with shot‑noise limit. Figure~\ref{fig:vqe}b shows the corresponding quantile distributions at different optimization steps. At the end of optimization, at least 25\% of the trajectories achieve an error below 0.01 for the energy density and, in the best cases, the optimization reaches the intrinsic ansatz limit.

While it remains challenging to make quantitative predictions for the optimization cost of MERA circuits in 2D systems, these 1D results suggest that, with appropriate optimizer choices and warm-up strategies, the number of quantum states required for optimization is comparable to the shot count needed to reach a desired target accuracy.

\section{Simulating XXZ chain by MERA circuits}\label{sec:XXZcircuits}

While we primarily demonstrate our approach using the Ising model, the framework is broadly applicable. To illustrate this, we apply it to the 1D XXZ model
\begin{equation}
    \hat{H}=\sum_i (\hat{S}^x_i\hat{S}^x_{i+1} + \hat{S}^y_i\hat{S}^y_{i+1} + \Delta \hat{S}^z_i\hat{S}^z_{i+1}).
\end{equation}
The XXZ spin chain exhibits a quantum phase transition at the isotropic point $\Delta = 1.0$, separating a gapless critical XY (Luttinger‑liquid) phase from a gapped N\'eel antiferromagnetic phase. This transition is of the Berezinskii-Kosterlitz-Thouless (BKT) type, and it is less conventional than the Ising case. At $\Delta=-1$ the system transitions to a (trivial) ferromagnetic phase.

Accurately describing more complex models requires larger MERA bond dimensions. We numerically simulate the XXZ model by designing MERA quantum circuits where we assign two qubits to each bond; this yields a bond dimension of $\chi = 4$, doubling the qubit resources compared to the Ising case. To implement the enlarged tensors efficiently, we impose an internal substructure comprising brick-wall steps of two-qubit blocks, following our previous work~\cite{Miao2023-5}. Each generic two-qubit block is parameterized by the unitary operator $U = [R_z(\theta_1) \otimes R_z(\theta_2)] R_{zz}(\theta_3) R_{yy}(\theta_4) R_{xx}(\theta_5)$~\footnote{To maximize compatibility with our specific gate parameterization, we apply a local basis transformation (alternating $\hat{X}$ flips on the lattice) that maps the standard XXZ Hamiltonian to $\hat{H}=\sum_i (\hat{S}^x_i\hat{S}^x_{i+1} - \hat{S}^y_i\hat{S}^y_{i+1} - \Delta \hat{S}^z_i\hat{S}^z_{i+1})$.}. Using MERA circuits with three steps per tensor and up to five layers, we obtain well‑behaved order parameters and entanglement entropy scaling in the XXZ model, as shown in Fig.~\ref{fig:XXZ}. In panel (a), the MERA circuit reproduces the sharp changes in order parameters near the phase transitions, analogous to Fig.~\ref{fig:local}a. More importantly, Fig.~\ref{fig:XXZ}b can resolve logarithmic entanglement scaling at $\Delta = 1.0$ and area‑law scaling at $\Delta = 1.7$. Although the resource requirements of this XXZ implementation (24 qubits and 540 native two‑qubit gates) exceed the capabilities of our current experimental setup, they are accessible in other state-of-the-art quantum hardware. For example, circuits with more than 2,000 two‑qubit gates and 56 qubits have been demonstrated on the Quantinuum H2 system (Ref.~\cite{haghshenas2025_03}).

\begin{figure*}
    \includegraphics[width=0.45\textwidth]{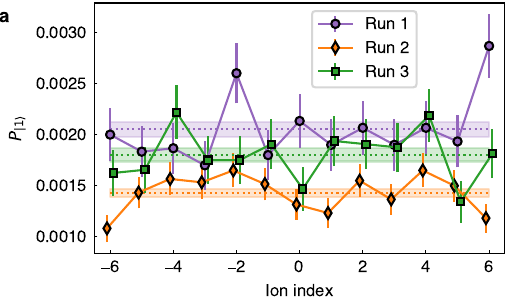}\hspace{0.3in}
    \includegraphics[width=0.45\textwidth]{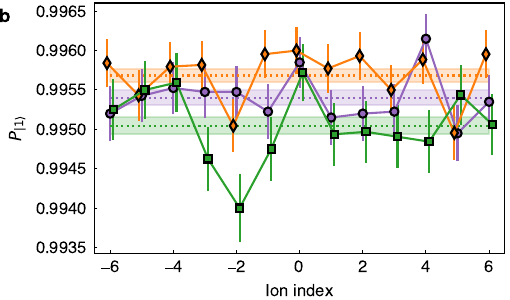}
    \caption{\label{fig:spam}\textbf{State preparation and measurement (SPAM) population as a function of the ion index.} \textbf{a} Dark-state ($|0\rangle$) SPAM experiments, with ion-averaged errors of 0.205(7)\%, 0.143(4)\%, and 0.180(7)\% for three runs. \textbf{b} Bright-state ($|1\rangle$) SPAM experiments, with ion-averaged errors of 0.460(9)\%, 0.432(9)\%, and 0.495(11)\%  for three runs. Error bars and shaded areas indicate $1\sigma$ statistical intervals.}
\end{figure*}

\section{System characterization and error model}\label{sec:system}
\subsection{State preparation and measurement (SPAM) error}\label{sec:SPAM}

\begin{figure*}
    \includegraphics[width=1\textwidth]{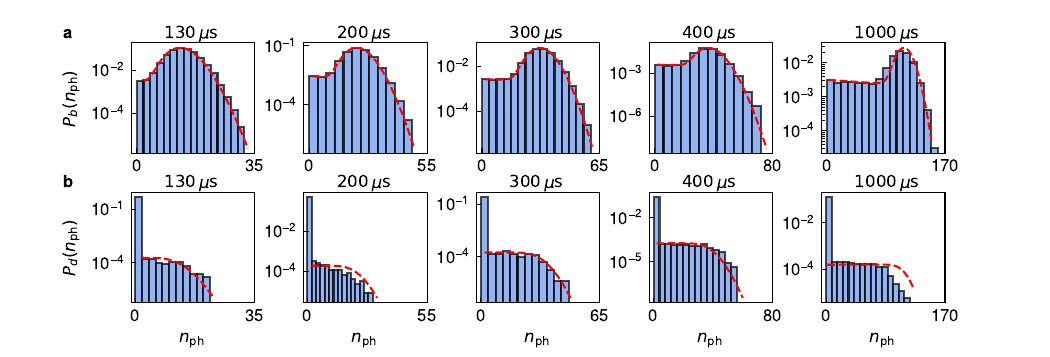}
    \caption{\label{fig:histogram_fit} Distribution of the detected photon counts $n_\text{ph}$ for different detection times for the bright state (\textbf{a}) and the dark state (\textbf{b}). The red dashed lines indicate a fit to the models from Eq.~\eqref{eq:nphb} and \eqref{eq:nphd}.}
\end{figure*}

\begin{figure}[t]
    \includegraphics[width=1\columnwidth]{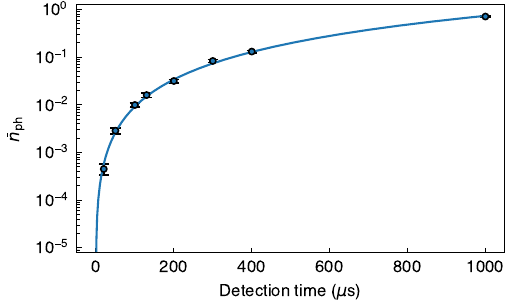}
    \caption{\label{fig:dark_n_avg fit}
    \textbf{Off-resonant pumping from the dark to the bright state during detection.} The average number $\bar{n}_\text{ph}$ of collected photons when preparing the ion in the $|0\rangle$ is plotted as a function of detection time $\tau$. The error bars indicate $1\sigma$ confidence interval with 60,000 experimental shots.}
\end{figure}

SPAM errors contribute significantly to our error budget. To quantify the probability $p_d$ of SPAM errors for the dark state ($|0\rangle$), we performed fluorescence detection after initializing all qubits via the usual  optical pumping, yielding an average measured error of 0.16\% (see Fig.~\ref{fig:spam}a for run-by-run results on different experimental days). 

To measure the probability $p_b$ of SPAM errors for the bright state ($|1\rangle$), we applied $R_y(\pi)$ gates following the optical pumping and then performed fluorescence detection, yielding an average measured error of 0.45\% (see Fig.~\ref{fig:spam}b). 
We checked the rotation error of the $R_y(\pi)$ gates by applying nine consecutive $R_y(\pi)$-gates, yielding a population error of $0.56^{+0.17}_{-0.12}\%$ as determined by the binomial proportionate 95\% confidence interval using the Clopper-Pearson method. The population error resulting from single-qubit gates should scale superlinearly with the number $N$ of gates, corresponding to a mixture of uncorrelated and correlated noise. Our results imply a population error below $10^{-3}$ per $\pi$ gate, which is well below the determined average bright-state SPAM error of 0.45\%.

We expect our SPAM errors to be dominated by the measurement error, with an additional preparation error for the $|0\rangle$ state.
The measurement error for the $^{171}$Yb$^+$ hyperfine qubits arises from optical pumping between the hyperfine states during the optical detection pulse \cite{crain_high-speed_2019,roman_coherent_2020}. We fit the measured photon count statistics at various detection time $\tau$ to an optical pumping model as discussed below to determine the state preparation error and the measurement errors.

The distributions of detection counts for a single ion prepared in the bright state for different detection times are shown in Fig.~\ref{fig:histogram_fit}a. Ideally, the ion would cycle between the $\ket{S, F=1}$ and $\ket{P, F=0}$ manifolds, resulting in a Poisson distribution of the recorded photon counts. However, the bright state can be off-resonantly pumped to the dark hyperfine ground state via the $\ket{S, F=1}\to\ket{P, F=1}$ transition detuned by 2.1 GHz with a rate $R_b$ that is about $5\times10^4$ times lower than the bright-state photon scattering rate $\Gamma_{\text{sc}}$. Neglecting the even slower dark-to-bright pumping process, the predicted distribution of the bright-state photon counts are
\begin{equation}\label{eq:nphb}
    P_b(n) = \mathrm{e}^{-R_b \tau}\mathcal{P}(n;\eta \Gamma_{\text{sc}} \tau) + \int_0^\tau \!\mathrm{e}^{-R_b t} \mathcal{P}(n;\eta \Gamma_{\text{sc}} t)\, R_b \ \mathrm{d}t,
\end{equation}
where $\mathcal{P}(n;\bar{n})=e^{-\bar{n}}\bar{n}^n/n!$ is the Poisson distribution with mean $\bar{n}$, and $\eta$ is the detection efficiency. The red dashed curves in Fig.~\ref{fig:histogram_fit}a show fits of the obtained count data to this model with $\eta \Gamma_{\text{sc}} \tau$ and $R_b$ as the fitting parameters. We average the fitted $R_b$ values with weights inversely proportional to the  fitted variances to obtain $R_b=3.0(1)\times10^2$/s. 

The detection count statistics for a single ion prepared in the dark $\ket{S, F=0}$ state are shown in Fig.~\ref{fig:histogram_fit}b. The dark state can be off-resonantly pumped to the bright manifold via the $14.7$-GHz detuned $\ket{S, F=0}\to\ket{P, F=1}$ transition, with a rate $R_d$ that is about $10^6$ times lower than $\Gamma_{\text{sc}}$. Neglecting the reverse pumping process yields the predicted dark-state count distribution 
\begin{equation}\label{eq:nphd}
     P_d(n) = \int_0^\tau \mathcal{P}(n;\eta\Gamma_{\text{sc}} (\tau-t))\, e^{-R_d t} R_d\ \mathrm{d}t.
\end{equation}
Fitting this model to the count data for photon number $n_\text{ph}>1$ in Fig.~\ref{fig:histogram_fit}b yields $R_d=18(1)$/s. The obtained ratio of $R_b$ and $R_d$ is close to the theoretical prediction in Ref.~\cite{noek_integration_2013}. 

To distinguish the preparation error from the measurement error, we fitted the average dark-state photon number as a function of detection time $\tau$ (Fig.~\ref{fig:dark_n_avg fit}) to the rate equation model from Ref.~\cite{noek_integration_2013}:
\begin{equation}\label{avgnph}
    \bar{n}_{\text{ph}} = \int_0^\tau \eta\, \Gamma_{\text{sc}} \left(P_{d,\infty} - (p - P_{b,\infty})\, \mathrm{e}^{-(R_b+R_d) t}\right) \mathrm{d}t ,
\end{equation}
where $p$ is the dark-state preparation error probability. Here, $P_{b,\infty}=(R_b/R_d + 1)^{-1}$ and $P_{d,\infty}=1-P_{b,\infty}$ are the equilibrium bright and dark-state populations, respectively.
The fitted value $p = 10(6)\times10^{-5}$ confirms that our SPAM error is dominated by the measurement errors, as $p\ll p_d,p_b$.

To model state-preparation error in the numerical simulations, we flip each qubit with probability $p$. To simulate the detection error, after applying the circuit and measuring in the computational basis $\{|0\rangle,|1\rangle\}$, we randomly flip $|0\rangle\rightarrow|1\rangle$ with probability $p_d - p$ and $|1\rangle\rightarrow|0\rangle$ with probability $p_b - p$.

\begin{figure}[t]
    \includegraphics[width=\columnwidth]{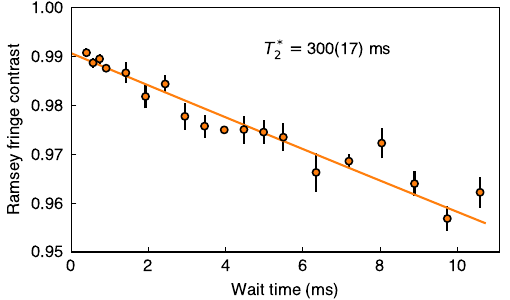}
    \caption{\label{fig:dephasing} \textbf{The average qubit Ramsey fringe contrast as a function of the idle time $\tau_\text{idle}$ from a sample dataset.} The orange line represents the fit to the model $A\,\mathrm{e}^{-\tau_\text{idle}/T^*_2}$ yielding $T_2^* = 300(17)$ ms and $A=0.991(1)$. For the experimental data, error bars indicate 1$\sigma$ fitting uncertainties. The deviation of the fitted value of $A$ from $1$ mainly reflects SPAM errors, with additional contributions potentially arising from residual single-qubit gate errors and qubit addressing and detection crosstalk. 
}
\end{figure}

\subsection{Idling noise}\label{sec:idle}
We characterized the dephasing of idle qubits using a Ramsey pulse sequence. After applying $R_y (\frac{\pi}{2})$ gates to all 13 qubits, we simulated idling by turning off the individual-addressing Raman beams while keeping the global beam on with the same settings as during a gate sequence. After a variable idle time $\tau_\text{idle}$, we completed the Ramsey sequence with $R_z (\phi)$ and $R_y (\frac{\pi}{2})$ gates and recorded the $|1\rangle$ state population as a function of $\phi$. We obtained the fitted ion-averaged contrast $C$ from the resulting fringes and modeled it as $A\,\mathrm{e}^{-\tau_\text{idle}/T^*_2}$. 
Figure~\ref{fig:dephasing} shows a representative fit from a sample dataset, yielding $A=0.991(1)$ and $T_2^*=300(17)$ ms. 
From several datasets collected on different experimental days, the fitted $T_2^*$ ranges from 260 ms to 370 ms, with an average of 290 ms. 
We attribute the measured dephasing to acoustic noise, mainly from air cooling fans, in our phase-sensitive Raman interferometer setup.
We expect that $T_2^*$ times of several seconds can be achieved using phase-insensitive schemes~\cite{Pino2021-592} that are compatible with our experimental setup.

In our simulations, we model the effects of dephasing on the qubit density matrix $\hat{\rho}$ over the time interval $t$ via the quantum channel
\begin{equation}\label{pZ}
    \mathcal{E}_Z(\hat{\rho}) = (1-p_Z)\hat{\rho} + p_Z \hat{Z} \hat{\rho} \hat{Z},
\end{equation}
with the $Z$-flip probability $p_Z = t/(2 T_2^*)$.

\begin{figure}
    \includegraphics[width=1\columnwidth]{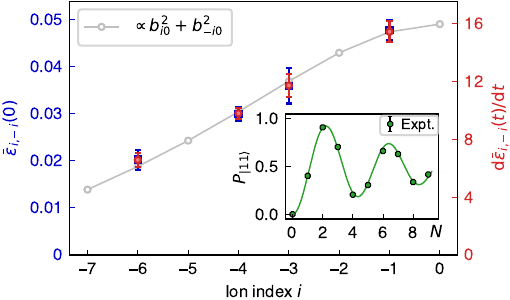}
    \caption{\label{fig:axial}
    \textbf{Gate errors due to axial motion.}
    The fitted initial decay parameter $\bar{\varepsilon}_{i,-i}(0)$ (blue) and its increase rate $\mathrm{d}\bar{\varepsilon}_{i,-i}(t)/\mathrm{d}t$ (red) are shown for $X\!X$ gates acting on ions $i$ and $-i$. The gray line is proportional to the sum of squares $b_{i0}^2 + b_{-i0}^2$ of the participation factors for the addressed ions in the lowest axial mode of the ion chain. Error bars indicate 1$\sigma$ fit uncertainties. The inset shows the $|1,1\rangle$-state population for ions $(-3,3)$ as a function of the gate number $N$ after a wait time of $t\approx 10$ ms, with a fit according to Eq.~\eqref{eq:theta}. The 1$\sigma$ statistical uncertainties (error bars) are smaller than the marker size.}
\end{figure}

\subsection{Gate errors}\label{sec:gateError}

While we actively calibrated the rotation angles, blue-and-red sideband imbalances, and Stark shifts of two-qubit gates before running circuits, residual two-qubit errors still remain and are discussed below.

\subsubsection{Axial motion}\label{sec:axial}
Axial motion of our 15-ion chain, dominated by the lowest-frequency axial mode (mode 0) with angular frequency $\omega_0 = 2\pi\times241.8$ kHz, is a leading error source in our platform. This motion misaligns the ions relative to the tightly-focused addressing beams, causing gate errors~\cite{Cetina2022-3}. These errors increase over time as the ions' motion is driven by noisy electric fields.

We quantified the addressing error after a wait time $t$ following laser cooling by applying a variable number $N$ of $X\!X(\frac{\pi}{2})$ gates of total duration $\Delta t$ to ions $i$ and $j$ and measuring the mean population $P_{|11\rangle}$ of the $|1,1\rangle$ state. The recorded populations $P_{|11\rangle}$ as a function of $N$ for $(i,j)=(3,-3)$ are shown in the inset of Fig.~\ref{fig:axial}.
Uncertainty in the ions' axial motion translates into a variance in the total $X\!X$ rotation angle, leading to damping of $P_{|11\rangle}$ oscillations. 

The observed damping is governed by the statistics of the phonon number of mode 0 during gate application~\cite{Cetina2022-3}. We model the evolution of the phonon number $n_t$ at time $t$ as a biased random walk. Since the energy of mode 0 following laser cooling corresponds to hundreds of phonons, and the average heating per $X\!X$ gate corresponds to more than 20 phonons, we coarse-grain $n_t$ on intervals of size $\delta n$ with $1\leq\delta n\ll2n_t$ and model the random walk as
\begin{equation}\label{eq:RW}
    n_{t+\delta t} := 
    \begin{cases}
        n_t + \delta n & \text{ with prob. } \frac{n_t + \delta n}{\delta n} \frac{\dot{\bar{n}}\delta t}{\delta n}\\
        n_t - \delta n & \text{ with prob. } \frac{n_t}{\delta n} \frac{\dot{\bar{n}}\delta t}{\delta n}\\
        n_t & \text{ otherwise,}
    \end{cases}
\end{equation}
where $\dot{\bar{n}}$ is the heating rate of mode 0, and the time step $\delta t$ is chosen so that the change of $n_t / \delta n$ remains small ($\frac{2n_t + \delta n}{\delta n} \frac{\dot{\bar{n}}\delta t}{\delta n}\leq1$). 

When initialized with a Boltzmann distribution with mean $\bar{n}_0 \gg 1$, $n_t$ remains exponentially distributed with mean $\bar{n}_t = \bar{n}_0 + \dot{\bar{n}} t$. In this thermal limit, the time-averaged phonon number $n_t^\text{avg}:=\frac{1}{\Delta t}\int_t^{t+\Delta t} n_{t^\prime} \,\mathrm{d}t^\prime$ between time $t$ and $t+\Delta t$ has the mean $\mu_t = \bar{n}_{t+\Delta t/2} = \bar{n}_t  + \dot{\bar{n}}\frac{\Delta t}{2}$ and the variance $\sigma_t^2 = \mu_t^2 - \frac{\bar{n}_t\dot{\bar{n}}\Delta t}{3}+\frac{\dot{\bar{n}}^2 \Delta t^2}{6}$. The distribution of $n_t^\text{avg}$ can be well approximated by a Gamma distribution with the corresponding mean and variance.

If $N$ $X\!X(\theta)$ gates are applied between time $t$ and $t+\Delta t$, neglecting other errors and assuming the above Gamma distribution, we obtain the population 
\begin{equation}\label{eq:theta}
    P_{|11\rangle}(t) = \frac{1-C\cos(N \theta-\phi)}{2},
\end{equation}
of the $|1,1\rangle$ state with phase shift $\phi = \alpha \arctan(\alpha N \theta \bar{\varepsilon}_{i,j}(t + \frac{\Delta t}{2}))$ and contrast 
$C = (1 + [\alpha N \theta \bar{\varepsilon}_{i,j}(t + \frac{\Delta t}{2}) ]^2)^{-{\alpha}/{2}}$, where $\alpha = {\mu_t^2}/{\sigma_t^2}$ and $\bar{\varepsilon}_{i,j}(t)=\bar{\varepsilon}_i(t) + \bar{\varepsilon}_j(t)$. 
The mean dimensionless decay parameter $\bar{\varepsilon}_{i}(t)$ during the gate sequence is
\begin{equation}\label{eq:n2epsilon}
    \bar{\varepsilon}_{i}(t) = \frac{\hbar b_{i0}^2}{m\omega_0 w^2} \bar{n}_t,
\end{equation}
where $b_{i0}$ is the participation factor of ion $i$ in mode 0. Here, we assume that the ion motion remains much smaller than our Gaussian-shaped beams. We determine the individual-beam waist $w$ by fitting the dependence of the single-qubit Rabi frequency $\Omega$ on  the axial ion position $x$ to $\Omega_0 e^{-(x-x_0)^2/w^2}$, yielding $w=646(12)$ nm.
In the limit of negligible heating during the gate sequence, the prediction~\eqref{eq:theta} reduces to Eq.~(4) in~\cite{Cetina2022-3}.

For different ion pairs $(i,j=-i)$, we fit the measured mean $|1,1\rangle$-state population as a function of the gate number $N$ and wait time $t$ to the model \eqref{eq:theta}. The fitted dimensionless effective decay parameter $\bar{\varepsilon}_{i,j}(0)$ at $t=0$ and its rate of increase $\mathrm{d}{\bar{\varepsilon}}_{i,j}/\mathrm{d}t$ are shown as functions of $i$ in Fig.~\ref{fig:axial}. Both parameters scale with $b_{i0}^2 + b_{j0}^2$, indicating that addressing errors are dominated by excitations of the lowest axial mode of the ion chain. The fitted initial phonon number in this mode is $\bar{n}_0=409(4)$ and the heating rate is $\dot{\bar{n}}=133(1)$ phonons/ms. 

To model gate rotation errors for arbitrary single- and two-qubit gates, in each shot of the noisy simulations, we sample the initial phonon number $n_0$ from an exponential distribution with mean $\bar{n}_0$. We then evolve $n_t$ using the random walk model~\eqref{eq:RW}. For each native gate in the circuit, we compute the time-averaged phonon number $n_t^\text{avg}$. In the noise model, the phonon number $n_t^\text{avg}$ changes single-qubit gate rotations as $R(\theta)\rightarrow R(\theta(1- n_t^\text{avg}\bar{\varepsilon}_i/\bar{n}_t))$ and two-qubit gate rotations as $X\!X (\theta)\rightarrow X\!X(\theta(1- n_t^\text{avg}\bar{\varepsilon}_{i,j}/\bar{n}_t)$.

\begin{figure}[t]
    \includegraphics[width=0.95\columnwidth]{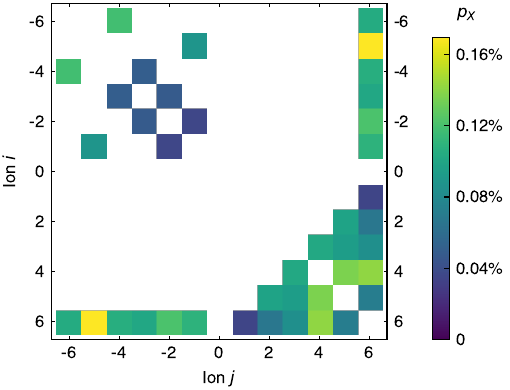}
    \caption{\label{fig:px} 
    \textbf{Pauli $X$ error per fully-entangling $X\!X$ gate.} The Pauli $X$ error $p_{X,i,j}(\pi/2)$ for gates acting on ions $i$ and $j$ is calculated using Eq.~\eqref{eq:pXij}.
    We only show error data for gates that are actually utilized in this work.}
\end{figure}

\subsubsection{$X$-errors}\label{sec:Xerror}
We describe $X$-errors during an $X\!X(\theta)$ gate on ion pair $(i,j)$ using a bit-flip channel acting on the density matrix $\hat{\rho}$ of ion $i$ as
\begin{equation}\label{eq:EX}
    \mathcal{E}_{X}(\hat{\rho}) = (1-p_{X})\hat{\rho} + p_{X} \hat{X} \hat{\rho} \hat{X},
\end{equation}
where $p_{X}\equiv p_{X,i,j}(\theta)$ is the $X$-flip probability per gate.

$X$-errors can be caused by residual spin-motion entanglement at the end of M\o{}lmer-S\o{}rensen gates. They can be minimized by carefully choosing the motional detuning, gate duration, and envelope of the amplitude-modulated gate waveforms~\cite{Huang2024-10}. In-between experiment runs, we stabilized frequencies of the radial ion-motion modes via blue-sideband Ramsey spectroscopy. Each of the central 11 qubits was used to query one of 11 radial modes that are addressed by the entangling gate waveforms. The obtained spectroscopical data was used to automatically adjust the quadratic and quartic terms in the static axial potential of the trap and a common motional offset frequency for all gates.

After applying $N$ consecutive $X\!X(\frac{\pi}{2})$ gates to the $|0,0\rangle$ state on ions $i$ and $j$, we measured the probability of $X$-flip errors by tracking the population $P_{i,j}(N)$ of the $|0,1\rangle$ and $|1,0\rangle$ states (parity leakage). We distinguished $X$-flip errors from coherent $X\!X$ gate crosstalk by focusing on the shots where only the target qubits are flipped.
In the noise model \eqref{eq:EX}, we employ the $X$-flip error probabilities
\begin{equation}\label{eq:pXij}
    p_{X,i,j}(\theta)=\frac{|\theta|}{\pi/2}\frac{P_{i,j}(N)}{2N},
\end{equation}  
which are summarized in Fig.~\ref{fig:px}.

\begin{figure*}[t]
    \includegraphics[width=0.95\columnwidth]{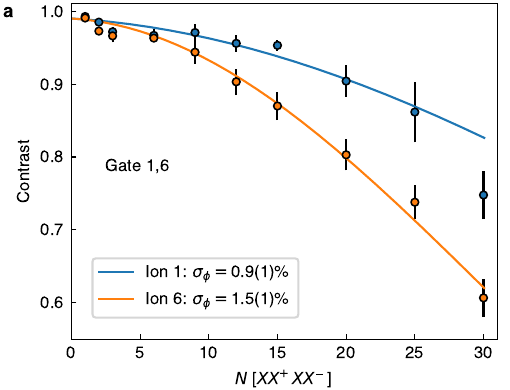}\hspace{0.3in}
    \includegraphics[width=0.95\columnwidth]{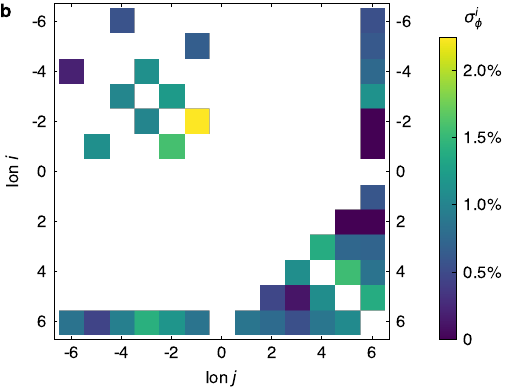}
    \caption{\label{fig:SS} 
    \textbf{Effect of Stark shifts on $X\!X$ gates.} \textbf{a} Ramsey fringe contrast as a function of the number of $(X\!X^+,X\!X^-)$ gate pairs inside the Ramsey sequence for a sample gate. Error bars indicate 1$\sigma$ fit uncertainties. Solid lines indicate fits to a Gaussian function with zero mean and standard deviation $N\sigma^i_\phi$. \textbf{b} Standard deviation $\sigma^i_\phi$ of the random Gaussian-distributed phase shift $\phi_i$ on ion $i$ per fully-entangling $X\!X$ gate acting on ion pair $(i,j)$. Error data are shown only for gates utilized in this work.} 
\end{figure*}

\subsubsection{$Z$-errors}\label{sec:Zerror}
To measure Stark shifts during a specific $X\!X$ gate, we insert $N$ pairs of $X\!X(\frac{\pi}{2}) - X\!X(-\frac{\pi}{2})$ gates into a Ramsey pulse sequence consisting of initial $R_y(\frac{\pi}{2})$ and final $R_z(\phi) - R_y(\frac{\pi}{2})$ pulses applied to each targeted ion. After adjusting the qubit frequencies during the gate to compensate for the measured shifts, we use the same sequence to estimate any residual gate $Z$-errors.

The fitted contrast of the obtained Ramsey fringes for a sample gate is shown in Fig.~\ref{fig:SS}a. The observed dependence of the contrast on $N$ is roughly quadratic, indicating that the phase shift error accumulates coherently as the number of gates increases. 
We model the observed decoherence by correlated fluctuations in the gate-induced Stark shift, which can arise from fluctuations in the amplitude balance of the red and blue tones of the M\o{}lmer-S\o{}rensen gate due to air turbulence in the Raman beam setup. Since our Ramsey sequence prepares the qubits in the $X$-basis, we do not expect gate $X$-errors to affect the measured contrast.
We assume that the phase shift $\phi_i$ induced on ion $i$ by a fully-entangling $X\!X$ gate during each shot follows a Gaussian distribution with zero mean and standard deviation $\sigma_\phi^{i}$. This noise model results in a fringe contrast proportional to $\mathrm{e}^{-(N\sigma_{\phi}^i)^2/2}$ to first order. Figure~\ref{fig:SS} shows fits of this model to the measured contrast data and the fitted values of $\sigma_\phi^i$ for all fully-entangling $X\!X$ gates on the ion pairs used in this work.

In the noise-model simulations, for each shot,  we sample random numbers $s$ from the standard normal distribution $\mathcal{N}(0,1)$. When executing quantum circuits, we replace the ideal $X\!X(\theta)$ gate by 
\begin{equation}
    \mathrm{e}^{-\frac{\mathrm{i}}{2} \left( \theta \hat{X}_i\hat{X}_j + \phi_i\hat{Z}_i + \phi_j\hat{Z}_j\right)},\quad\text{where}\ \ 
    \phi_i=s\frac{\sigma^i_\phi |\theta|}{\pi/2}
\end{equation}
is a rescaled phase shift on ion $i$. 

\section{Breakdown of error contributions}\label{sec:breakdown}

\begin{figure*}[t]
    \centering
    \includegraphics[width=0.9\textwidth]{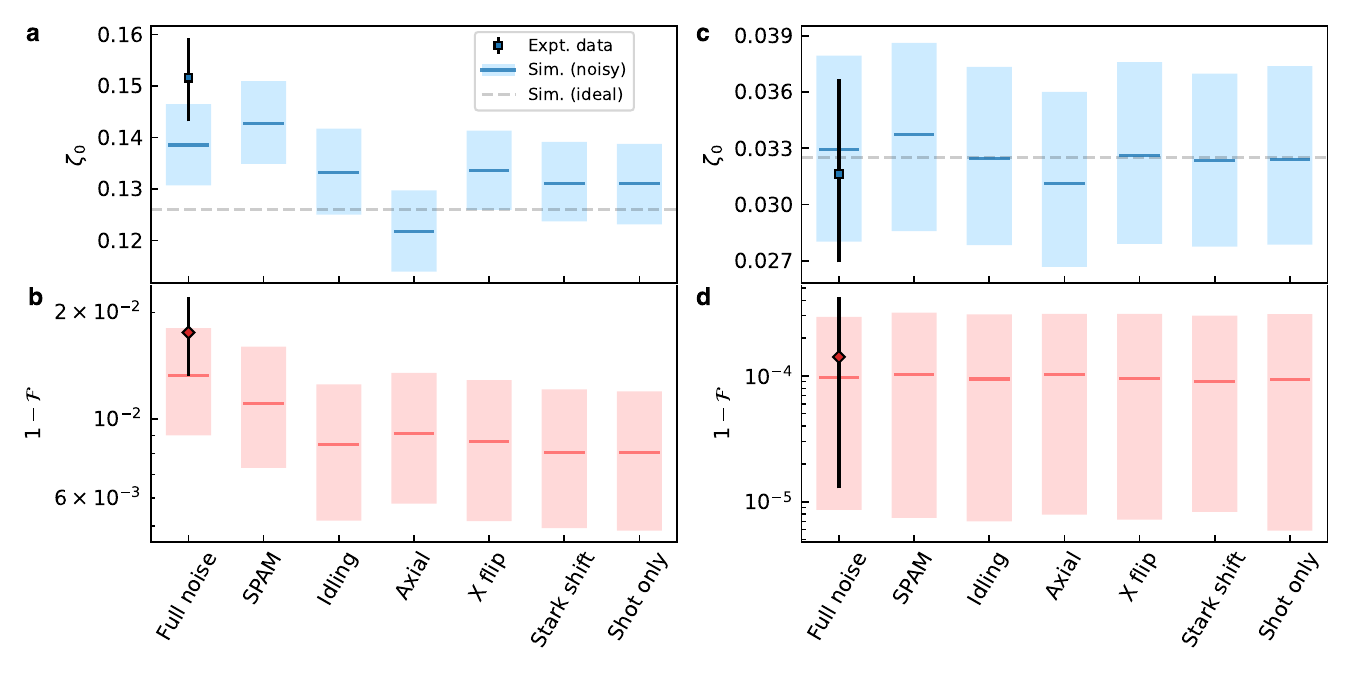}
    \caption{\label{fig:errorDetails}\textbf{Breakdown of error contributions.} The top row (\textbf{a}, \textbf{c}) shows the first eigenvalues $\zeta_0$ of the entanglement Hamiltonian, and the bottom row (\textbf{b}, \textbf{d}) shows corresponding state infidelities $1-\mathcal{F}$. The left column (\textbf{a}, \textbf{b}) presents results at the critical point $g=1$ with $T=4$, while the right column (\textbf{c}, \textbf{d}) corresponds to the gapped phase at $g=1.5$ with $T=1$. Experimental data points with error bars are reproduced from Fig.~\ref{fig:EH} in the main text. The dashed lines indicate the ideal simulation results from Fig.~\ref{fig:EH}. Noisy simulation results are shown as shaded regions representing the 95\% quantiles, with solid colored lines indicating the mean values.
    In the ``full noise'' case, we reproduce the full noisy simulation including all error sources from Fig.~\ref{fig:EH}. In the ``shot only'' case, all errors are disabled except for shot noise (the statistical uncertainty due to the finite number of measurement shots). In the remaining cases, the simulations include both the shot noise and the specified individual error source. We run 1000 independent noisy simulation trials for each case. Each trail uses the same number of measurement shots as the experiment: 1,500 shots per basis per trial in the left column, and 8,000 shots per basis per trial in the right column.
    }
\end{figure*}

To better understand our experimental results, we analyze the breakdown of error contributions through numerical simulation. As shown in Fig.~\ref{fig:errorDetails}, we selected two representative cases from Fig.~\ref{fig:EH} of the main text, one with the highest reported infidelity $1-\mathcal{F}$ ($\sim10^{-2}$) and one with the lowest ($\sim10^{-4}$).

The case of highest reported infidelity corresponds to a 4-qubit tomography result on a 4-layer MERA circuit at the critical point $g = 1$. The circuit involves 9 qubits and 21 $X\!X$ gates with rotation amplitudes ranging from $0.06\pi$ to $0.18\pi$. We performed 1,500 measurement shots per basis over a total of $3^4=81$ measurement bases. Based on our error model and the calibrated parameters provided in 
App.~\ref{sec:system}, we identify several key sources of error. First, there is the inevitable statistical uncertainty
in the measured density matrix elements on the order of $\sim10^{-2}$ from the finite number of measurement shots. Second, aggregated SPAM errors for the 4-qubit tomography are on the order of $\sim10^{-2}$. The dominant gate error arises from rotation-angle fluctuations caused by axial motion of the ion chain, which, for fully-entangling gates, typically leads to errors on the order of $10^{-2}$.
After accounting for the error rescaling due to smaller rotation angles in the actual MERA gates and the accumulation of errors throughout the circuit, the overall gate-induced error is also expected to be in the $10^{-2}$ range. This estimate is confirmed in Fig.~\ref{fig:errorDetails}a. A small discrepancy remains between the experimental data and the full-noise simulation, which we attribute to system miscalibration over the long experimental runtime required for 4-qubit tomography, as well as unmodeled single-qubit gate errors. 

The case of lowest reported infidelity corresponds to a single-qubit tomography result on a single-layer MERA circuit in the gapped phase at $g=1.5$. The circuit involves only 2 qubits and a single $X\!X$ gate with a rotation amplitude of $0.095\pi$. We performed 8,000 measurement shots for each Pauli-$X$, -$Y$, and -$Z$ basis. In this case, the statistical uncertainty from finite sampling is on the order of $10^{-3}\sim 10^{-2}$, while the SPAM error for a single qubit is on the order of $10^{-3}$, and the error from the single gate is also on the order of $10^{-3}$. Hence, we expect the statistical uncertainty to dominate over other error sources, which is consistent with the simulation results shown in Fig.~\ref{fig:errorDetails}c.

We observe that the infidelity is of comparable magnitude to the errors for the high-infidelity case in Fig.~\ref{fig:errorDetails}b and at least an order of magnitude below the error scale for the low-infidelity case in Fig.~\ref{fig:errorDetails}d. We attribute this to the fact that the infidelity has different scaling behavior for rank-deficient and full-rank states: Consider the exact (noise-free) density operator $\hat{\rho}$ and the perturbed version $\hat{\sigma} = \hat{\rho} + \epsilon \hat{D}$ with noise level $\epsilon$ and $\mathrm{Tr}(\hat{D}) = 0$. For a full-rank state $\hat{\rho}$, the perturbed $\hat{\sigma}$ remains well-defined (positive semidefinite) for sufficiently small $\epsilon > 0$ as well as small $\epsilon < 0$. The infidelity $1 - \mathcal{F}(\hat{\rho}, \hat{\sigma}(\epsilon))$ is differentiable and minimal at $\epsilon = 0$. This implies that the linear term vanishes and $1 - \mathcal{F} = \mathcal{O}(\epsilon^2)$. This is the case in Fig.~\ref{fig:errorDetails}d, where the corresponding single-qubit reduced density matrix has full rank with eigenvalues $\{0.98, 2.2\times10^{-2}\}$. A linear scaling $1 - \mathcal{F} = \mathcal{O}(\epsilon)$ is only possible for rank-deficient states. For example, consider a pure state $\hat{\rho} = |\psi\rangle\langle\psi|$ and the perturbation $\hat{D} = -|\psi\rangle\langle\psi| + |\psi_\perp\rangle\langle\psi_\perp|$ with an orthogonal state $|\psi_\perp\rangle$ and $\epsilon \ge 0$. In this case, the infidelity is simply $\epsilon$. In the current experiments, the ideal reduced density matrix is indeed approximately rank-deficient at larger layer numbers $T$, and this is the case in Fig.~\ref{fig:errorDetails}b with eigenvalues $\{0.92,7.9\times10^{-2},3.5\times10^{-3},3.1\times10^{-4},2.7\times10^{-4},2.7\times10^{-5},\dotsc\}$.

\bibliography{ref}

\end{document}